\documentclass[12pt]{book}

\usepackage[dvips]{graphicx,color}
\usepackage{makeidx,tsukuba,amssymb}

\makeauthorindex
\makeindex

\begin{document}

\BookTitle{\itshape The 28th International Cosmic Ray Conference}
\CopyRight{\copyright 2003 by Universal Academy Press, Inc.}
\pagenumbering{arabic}

\chapter{
Report on the High Energy Phenomena
Sessions HE 2, HE 3.2-3.4: Neutrinos and Muons. Interactions, 
Particle Physics Aspects, Astro-Particle Physics
and Cosmology\\}

\author{%
Teresa Montaruli$^1$ \\
{\it (1) Universit\`a di Bari and INFN, Via Amendola 173, 70126 Bari, Italy, \\
e-mail: montaruli@ba.infn.it}\\
}

\section*{Abstract}
The results presented at the {\it 28th International Cosmic Ray 
Conference} on neutrino and muon physics are summarized.  
Neutrinos and muons provide a huge amount of information on particle 
interactions up to very high energies
and on fundamental particle properties. 
Results on neutrino oscillations in the atmospheric and solar 
$\nu$ sectors are summarized. 
Oscillations are well established in both sectors, 
and a more precise determination of oscillation parameters
is requested in the next future. 

Neutrino telescopes taking data and under construction presented
numerous results.
Neutrinos as probes of the Universe are hopefully going to open,
together with gravitational waves, a new era for Astrophysics.
Cosmology has entered the precision era and the Dark Matter quest is still 
an open problem. Direct and indirect searches are complementary 
approaches to the problem. 

The results presented at this conference confirm that
Astroparticle Physics and, in particular, Neutrino Physics
are leading fields in fundamental research. 

\section{Introduction}

The High Energy Phenomena Sessions HE 2.1-5 of this conference are 
devoted to: muon and neutrino experiments and calculations, 
neutrino telescopes and new projects. 
The Sessions HE 3.2-4 are dedicated to more exotic searches, such as 
dark matter and new particles, proton decay and cosmology, 
both from the theoretical and the experimental point of view. 
The total amount of 139 talks and posters cannot be entirely summarized 
here. In the attempt to provide an organic overview on the 
subjects of these sessions, some of the works are not mentioned. I do
apologize with the authors. Whenever possible experimental results and 
calculations are compared.

Sec.~\ref{sec:atmo} is devoted to neutrinos and muons of atmospheric origin.
In Sec.~\ref{sec:atmnu} the current status on our experimental
and theoretical knowledge of the atmospheric neutrino beam and
its impact on the oscillation scenario is discussed.
Super-Kamiokande (SK) has presented results on atmospheric neutrinos (HE 2.2)
and on $\nu_{\mu} \rightarrow \nu_{\tau}$ oscillations.  
In Sec.~\ref{sec:calc}, the status of atmospheric neutrino calculations
is summarized (HE 2.4), with particular
attention to the primary cosmic ray (CR) flux, a crucial input to
atmospheric cascade calculations.

Muon flux measurements (HE 2.1), currently, are mainly 
motivated to benchmark atmospheric shower development codes and hence
atmospheric $\nu$ calculations. In fact muons and neutrinos come 
from the same decay chains.  
In Sec~\ref{sec:muons} the results presented at the conference on muon
fluxes and charge ratios are summarized.

Sec.~\ref{sec:astro} is devoted to neutrinos of astrophysical origin,
not produced in the Earth atmosphere.
Among these, solar neutrinos are used to investigate neutrino
properties as well as the Sun itself.
In Sec.~\ref{sec:solar} results from SNO and SK (HE 2.2)
are summarized.
The solar neutrino experiment results together with the
KamLAND long baseline reactor experiment results are providing a convincing 
solution to the solar neutrino problem on $\nu_e$ disappearance. 

Currently solar neutrinos and events registered by Kamiokande and IMB [1] 
a few hours before the optical identification of 
SN1987A are the only astrophysical neutrinos
detected so far. In Sec.~\ref{sec:collapse} limits on supernova (SN) 
collapse are summarized. 
On the other hand, no neutrino of astrophysical origin has yet 
been detected above 
the GeV scale in the background of atmospheric neutrinos.
Nevertheless hopefully in a few years we will enter the neutrino astronomy 
era, thanks to the numerous efforts on the construction of huge neutrino 
telescopes to which Sec.~\ref{sec:nutelescopes} is devoted (HE 2.3). 
Results from various neutrino telescopes 
on point-like and diffuse sources are presented, 
including sensitivities expected for experiments
under construction and R\&D.

In Sec.~\ref{sec:dm} some of the results and sensitivities 
on dark matter searches through the detection of gammas, anti-protons,
positrons and neutrinos by satellites, ground-based arrays and neutrino 
telescopes will be summarized.

\section{Atmospheric Neutrinos and Muons}
\label{sec:atmo}

%
\subsection{Atmospheric Neutrinos and Super-Kamiokande}
\label{sec:atmnu}

Super-Kamiokande is an ultra-pure water Cherenkov detector, with a 
fiducial volume of 22.5 ktons and a 40\% phototube (PMT) coverage 
(11,146 51 cm Hamamatsu PMTs looking to the inner detector and 1885 
20 cm PMTs of the outer veto) [2]. 
A reaction chain in Nov. 2001 destroyed 50\% of the  PMTs.
This unfortunate event concluded the SK-I phase with a total amount of
1489 days of data taking (91.7 kt yr). 
In Dec. 2002 the reconstruction was completed 
and since Jan. 2003 the experiment has been taking new data and
the K2K neutrino beam has been on.

Even though no experiment has yet measured the oscillatory pattern,
which would unequivocally establish an oscillation phenomenon, the strength
of Super-Kamiokande result relies on the measurement of
topologies belonging to different energy ranges and 
of both electron and muon flavors.
The updated analysis presented at the conference introduced some refinements 
in the event reconstruction and in particle identification (on which the
systematic error is at the level of $1\%$ as estimated from Monte Carlo
studies, from a test of the KEK proton synchrotron [3], and confirmed
by the near 1 kton detector of K2K [4]). Also improvements
on the upward through-going muon selection have been adopted. 
Moreover, the Monte Carlo (MC) generator has been updated
using the calculation by Honda et al. [5], which adopts
the fit to primary CR measurements presented at ICRC2001 [6]. 
Also the quasi-elastic and $1\pi$ production cross 
sections were improved thanks to the K2K data.

Tab.\,1 summarizes the statistics and the values of the flavor ratio for 
the Sub-GeV fully contained (FC) sample 
($E_{vis}<1.33$ GeV) and the Multi-GeV 
(FC with $E_{vis}>1.33$ GeV) and partially contained (PC) sample 
($<E_{\nu}> \sim 10$ GeV). For the Multi-GeV sample 
the up/down asymmetry for $\mu$-like events 
deviates from zero (expected in the no oscillation case) 
by 9.5 $\sigma$: $A_{\mu-like}=
\left( \frac{N_{up}-N_{down}}{N_{up}+N_{down}} \right)=-0.289
\pm 0.028_{stat} \pm 0.004_{sys}$. On the other hand, for e-like events
$A_{e-like} = -0.020 \pm 0.043_{stat} \pm 0.005_{sys}$.
These measurements provide a robust evidence in favor of $\nu_{\mu}
\rightarrow \nu_{\tau}$ oscillations, as will be discussed 
in Sec.~\ref{sec:calc}. Muon neutrino charged current
(CC) interactions in the rock below the detector produce stopping
muons ($<E_\nu> \sim 10$ GeV) and upward through-going muons ($<E_\nu> \sim 
100$ GeV). The zenith angle distribution
of these events shows a distortion compatible to
oscillations. Tab.\,1 summarizes the statistics,
the measured and expected fluxes also for these samples.
\begin{table}[t]
\vskip -0.5 cm
\caption{Statistics and flavor ratios for the Sub-GeV and Multi-GeV
samples. Measured and expected fluxes (for no oscillations) are given 
for upward through-going $\mu$s and stopping $\mu$s }
\begin{center}
\begin{tabular}{cccc}
\hline
Sample & e-like  & $\mu$-like & $\frac{(\mu/e)_{data}}{(\mu/e)_{MC}}$ \\ \hline   
Sub-GeV & 3353 & 3227 &  $0.649 \pm 0.016_{stat} \pm 0.051_{sys}$ \\
Multi-GeV+PC & 746 & 1564 & $0.700 \pm^{0.032}_{0.030} \pm 0.083_{sys}$ 
\\ \hline \hline
Sample & Data & Measured flux & Expected Flux  [5] \\ 
       &   (1645.9 d)    & $10^{-13}$ cm$^{-2}$ s$^{-1}$ sr$^{-1}$ & 
$10^{-13}$ cm$^{-2}$ s$^{-1}$ sr$^{-1}$ \\ \hline
Stopping $\mu$s & 463 & $0.41 \pm 0.02_{stat} \pm 0.02_{sys}$&
$0.61 \pm 0.14$\\
Through-going $\mu$s & 1843 & $1.70 \pm 0.04_{stat} \pm 0.02_{sys}$ &
$1.57 \pm 0.35$ \\ 
\hline
\end{tabular} 
\end{center}
\end{table}

The SK-I updated analysis firstly presented at ICRC2003 [2] leads to 
a preliminary best fit $\Delta m^2$ value of $2 \cdot 10^{-3}$ eV$^2$, 
at maximal mixing, about $20\%$ smaller compared 
to the past result of $2.5 \cdot 10^{-3}$ eV$^2$ [7]. Allowed 
regions are compared in Fig.\,1, where the final result 
from MACRO [8] and Soudan 2 [9] are presented too. Also
the Kamiokande [7] result is shown: the lowering of SK
region increases the discrepancy between the SK and its precursor result
which could be clarified through a reanalysis of the Kamiokande data.
The 90\% c.l. region is $\sin^2 2\theta >0.9$ and        
$1.3 < \Delta m^2 < 3.0 \cdot 10^{-3}$ eV$^2$, while previously it was
$\sin^2 2\theta >0.9$ and $1.6 < \Delta m^2 < 3.9 \cdot 10^{-3}$ eV$^2$. 
The value of $\Delta m^2$ has an impact on the expectations of
Long Baseline Experiments, particularly on the
CNGS appearance experiments, OPERA and ICARUS. 
Considering a beam intensity larger than a factor 1.5 compared to the 
nominal beam, the CC $\nu_\tau$ 
events expected in ICARUS T3000 (2.35 kton active mass) in 5 yrs 
for the decay modes $\tau \rightarrow e \, , \rho$ 
vary from $\sim 7$ to 18 between 
$\Delta m^2 \sim 1.6-2.5 \cdot 10^{-3}$ eV$^2$ and maximal mixing [10],
while the background is $\sim 1$ event. 

Since one of the changes is related to the $\nu$ flux adopted by the 
experiment, the problem, addressed in Sec.~\ref{sec:calc}, 
on how well we do know the atmospheric $\nu$
flux in the various energy ranges arises, even though it is clear
that also the analysis refinements must have a relevant role. 

In the low energy range (FC events) SK investigated the 
geomagnetic field effects on $\nu$ fluxes 
through the measurement of the azimuth
distribution affected by the East-West anisotropy in the primary 
CR flux. The anisotropy results 
in larger (smaller) fluxes of $\nu$'s traveling toward East (West)  
[11]. This offers the possibility to validate 
calculations at $\lesssim 2$ GeV. SK is also sensitive to the finer
effect of an enhancement of the asymmetry for e-like events with
respect to $\mu$-like events that can be correctly taken into account
by 3-D MC calculations ($\nu$'s are not collinear to their
parents) [12].

SK can provide information on the oscillation channel: 
$\nu_{\mu} \rightarrow \nu_{sterile}$ oscillations are
disfavored at 99 \% c.l. through the non observation of
neutral current (NC) 
suppression using the $\pi^0$ sample which now is measured with a 
reduced systematic error (from 30\% to 9\%) thanks to the K2K
measurement of $R_{\pi^0} = \frac{\pi^0}{\mu}$ [13]. 
Moreover some indication on $\tau$ appearance can be extracted on a 
statistical basis [14].
\begin{figure}[t]
  \begin{center}
    \includegraphics[height=19.pc]{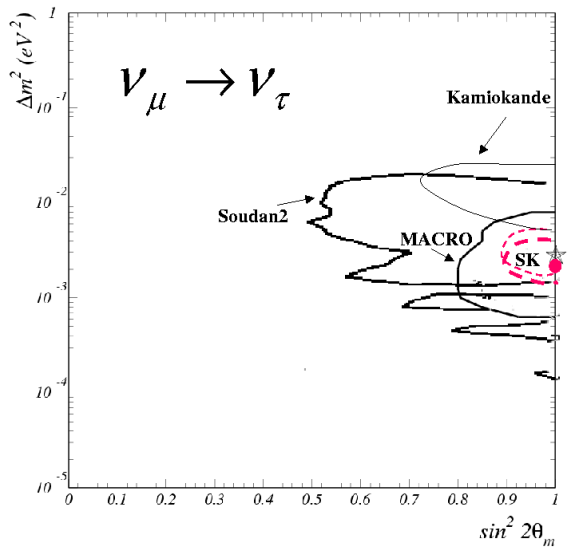} 
  \end{center}
  \vspace{-1.5pc}
  \caption{
The 90\% c.l. updated (dashed red line) and the previously (dotted thin
red line) allowed regions from the combined analysis of all samples for SK-I 
data. The Kamiokande (black solid thin line), MACRO (black solid line) [8], 
Soudan 2 [9] (blue 'irregular' solid line) are shown and indicated 
by names and arrows. 
SK (MACRO) updated best fit is indicated by a red circle (black star) 
at $\Delta m^2 = 2 \cdot 10^{-3}$ eV$^2$ ($2.3 \cdot 10^{-3}$ eV$^2$) and
maximal mixing.}
\end{figure}

\subsection{The Knowledge of the Atmospheric Neutrino Beam}
\label{sec:calc}

Exploiting different experimental techniques, SK, MACRO and Soudan 2 
provide a robust evidence 
in favor of $\nu_{\mu} \rightarrow \nu_{\tau}$ oscillations, 
since the flavor ratio, the up-down asymmetry and the zenith angle
distributions are
robust quantities against theoretical errors on the atmospheric
$\nu$ flux calculation, known at the level of $\lesssim 5\%$. 
As a matter of fact, at energies below about
2 GeV, when all muons decay in the atmosphere, 2 $\nu_{\mu}$ and 1 $\nu_e$ are
produced in $\pi$ and $\mu$ decay chains. Moreover,
far from regions where geomagnetic field effects are relevant, that is for 
$E_{\nu}\gtrsim 2$ GeV, 
the $\nu$ flux is up-down symmetric ($\Phi_{\nu} (E,\theta) = \Phi_{\nu}
(E, \pi-\theta)$, where $\theta$ is the zenith angle),
due to the atmosphere and Earth spherical symmetry. 
For higher energy events, i.e. the through-going muons,
the shape of the angular distribution obtained using different calculations
agrees at the level of $\sim 5\%$, 
where the remaining uncertainty is due to the knowledge of the competition 
between decay and interaction for pions and kaons, 
seasonal effects and variations in 
atmospheric profiles, and to the uncertainty in the spectral slope of the
primary flux.    
On the other hand, using the comparison between different
atmospheric neutrino calculations, it can be estimated that
the absolute normalization of fluxes is still affected by 15\%, 30\%
errors, 
respectively in the $\lesssim 10$ GeV and $\gtrsim 100$ GeV energy ranges.
This uncertainty is large even if calculations are
benchmarked against atmospheric muons at sea level and
in the atmosphere (see Sec~\ref{sec:muons}). 
The main errors come from the primary cosmic ray spectrum and
the hadronic interaction models. During the last years, authors of different
MC computations started a comparison work between interaction models
which led to code updates. For instance, the Bartol group 
presented at the conference
the updated generator TARGET [16, 17]. Other groups use
more sophisticated interaction and transport codes, such as
FLUKA [18], that has been extensively benchmarked on
accelerator data and muons (see for instance the comparison of
meson multiplicity distributions in $x_{lab}$ 
for generators used in CORSIKA [19]
with p-$^{9}$Be, p-$^{14}$N data),
the Japanese group updated the interaction 
model adopting DPMJET3 [5].  Wentz et al. adopted some of the
CORSIKA generators, among which DPMJET II.5 [20], while Liu et al. 
adapted a parametrized hadronic model published in 1989 [21]. 
It should be understood why Liu et al. calculation seems to 
predict a larger East-West asymmetry for $\mu$-like events than e-like,
contrarily to expectations (see Sec.~\ref{sec:atmnu}).
All of these codes are now 3-dimensional: 
the introduction 
of transverse momenta produces a considerable enhancement at the 
horizon for energies $\lesssim$ 2 GeV, first found in [18].
A review on atmospheric neutrino calculations is in [15].

The other important input in atmospheric neutrino calculations is
the primary CR spectrum. The uncertainties affecting the CR spectrum are
energy dependent. The recent fit presented at ICRC2001 [6]
was an attempt to unify this input between different calculations.
At energies $<200$ GeV/nucleon it relies on the recent measurements of 
the AMS-01 Space Shuttle flight [22] and of the balloon experiment
BESS 98 [23]: for protons they are in agreement 
within 5\%, while for He there is some discrepancy at the level of 15\%,
which at these energies reflects on a 3\% uncertainty on the all-nucleon 
flux [15]. 
Nevertheless, it remains to be understood the disagreement of 
the proton measurements by the CAPRICE balloon flights
(1994 [24] and 1998 [25]) which are about $20\%$ lower than AMS-01 
and BESS 98 results.

At high energy ($\gtrsim 10^4$ GeV/nucleon) the
JACEE [26] and RUNJOB [27,28] data have large errors leaving room
for large variations of the fitted spectral slope. 
While at energies $\lesssim 100$ GeV/nucleon, the ICRC2001 fit [6] 
changed by less than 10\% compared to the previous Bartol 96 fit used in
[29], at higher energies the ICRC2001 fit [6] is steeper  
($\sim E^{-2.74}$) than the Bartol 96 fit ($\sim E^{-2.71}$). 
In Fig.\,2 (on the left) the data and both fits are shown.
ATIC has presented at this conference preliminary data [30] 
from the long duration balloon flight 
that fill the gap in energy between previous experiments. 
These data are crucial to understand the slope of the spectrum and 
also if protons and He have the same slope, as it is discussed in [31]. 
Preliminary results indicate a $E^{-2.71}$ preference and that He spectrum has
almost the same slope of the proton one. A preference for
harder spectra for He (closer to JACEE result than to RUNJOB) 
is also found using EAS-TOP/MACRO and the analysis of
coincidences between the 2 experiments [32].

The effect on $\nu$ fluxes of changing the CR flux slope above
100 GeV/nucleon from $E^{-2.71}$ 
to $E^{-2.74}$ has been investigated by the Japanese [5] and FLUKA [18] groups.
Using both fits with the same interaction code, in the case of FLUKA, 
the ICRC2001 fit produces $\nu_\mu + \bar{\nu}_\mu$
fluxes (averaged over the entire hemisphere)
never larger by 10\% between 0.1-1 GeV, lower by 5\% at 10 GeV
and by 20-30\%  between 100 GeV-1 TeV than the flux
with the Bartol 96 fit.

One more indication on the fact that the ICRC2001 fit could be
too steep comes from SK and MACRO through-going $\mu$ data.
As a matter of fact, if the measured SK zenith angle distribution
of through-going muons is fitted letting the Honda et al. flux [5]
normalization free, 
the best fit is obtained for a normalization value
larger by $\sim 25\%$ than what predicted by [5] using the ICRC2001 fit. 
A similar result is obtained using MACRO upward through-going muons [8,18].
Nevertheless it is hard to understand the effect of the energy dependent
uncertainty on the normalization in the evaluation of the
best fit parameters. In fact the Super-Kamiokande fitting 
procedure contains various parameters
related to the normalization and the slope of the $\nu$ spectrum [2].
In Fig.\,2 (on the right) MACRO through-going
muons (821 events selected for oscillation studies) [8,18] are compared to the 
FLUKA fluxes with the ICRC2001 and 96 fits, 
the Bartol 96 flux and the Honda et al. flux [5] 
using the ICRC2001 fit. 
Clearly the data seem to prefer the fluxes using the 96 fit.
From the plot it is also noticeable the good
agreement between the calculations using two different interaction
models (FLUKA and DPMJET3) and the same CR spectrum (ICRC2001 fit).
\begin{figure}[t]
  \begin{center}
\begin{tabular}{cc}
    \includegraphics[height=19.5pc,width=17.5pc]{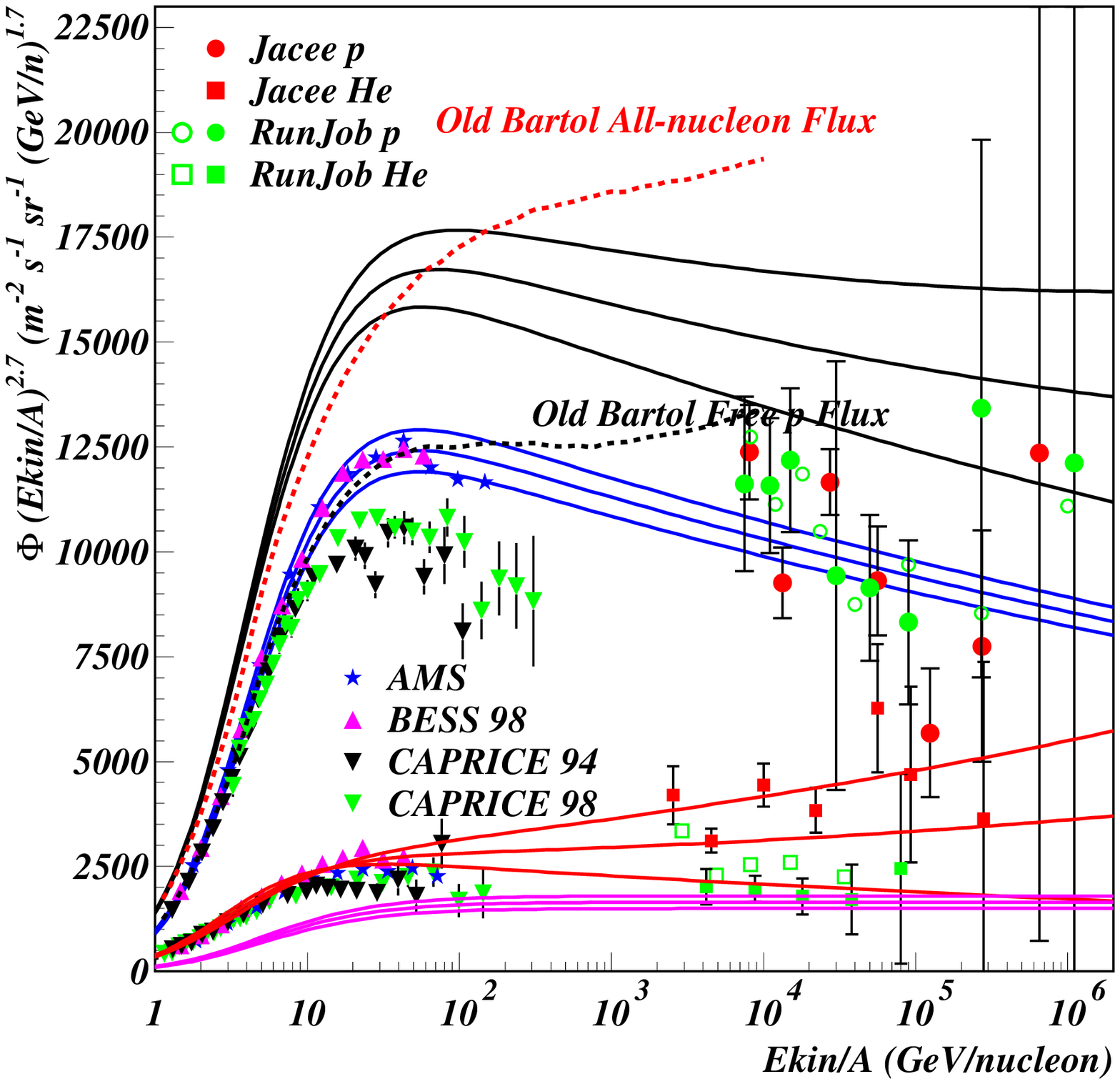} &
    \includegraphics[height=19.5pc,width=17.5pc]{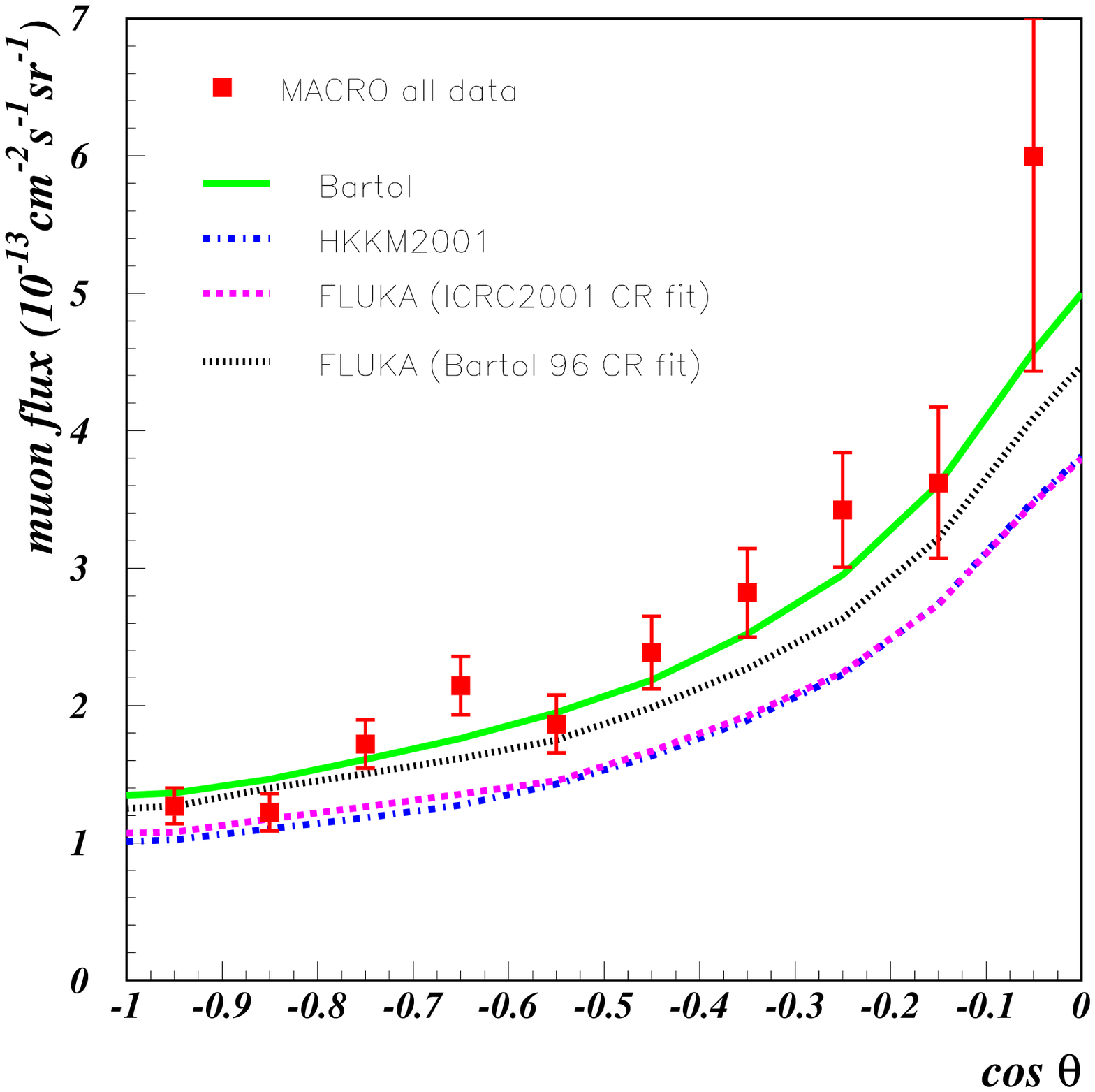} 
\end{tabular}
  \end{center}
  \vspace{-1.5pc}
  \caption{{\bf On the left:} The 3 upper black lines are the all nucleon primary 
spectrum resulting from the ICRC2001 fit [6]. The middle blue group of 
3 lines are the proton primary flux and the lower group the He flux.
The upper and lower lines of each group
are the maximum and minimum fluxes from the errors of the fit,
respectively. The dashed line is the all nucleon flux adopted 
by the Bartol group in 1996 [15]. The data points for protons and He
are from JACEE [26] and RUNJOB (open symbols: [27], full symbols: [28]),
AMS-01 [22], BESS 98 [23], CAPRICE 98 [25], CAPRICE 94 [24].
{\bf On the right:} upward through-going muon flux ($E_{\mu} > 1$ GeV)
measured by MACRO [8,18] (red symbols) compared to the Bartol 96 [29] (solid 
green line), to the FLUKA calculation using the Bartol 96 flux (black dotted 
line) and the ICRC2001 fit (pink dashed line), to the Honda et al. flux 
using the ICRC2001 fit [5] (blue dash-dotted line). Predicted curves
include oscillations with maximal mixing and $\Delta m^2 = 2.3 \cdot 10^{-3}$
eV$^2$.}
\end{figure}

\subsection{Muon Data at Sea Level and in the Atmosphere}
\label{sec:muons}

Many balloon and ground based experiments presented measurements 
of muon fluxes at sea level and in the atmosphere.
A summary of sea level data, including previous measurements is presented
in Fig.\,3 in terms of vertical muon flux and charge ratio $\mu^{+}/\mu^{-}$,
compared to the world average band estimated in [33]. At energies
$\gtrsim 5$ GeV muon data are not affected by solar modulation and geomagnetic 
cut-off effects, while for lower energies these effects must be considered,
as shown for instance by the comparison between the data taken at Ft. Sumner 
by CAPRICE 97 [34] and BESS-01 [35]. 

Preliminary results on atmospheric muons from LEP experiments have been 
presented
at the conference. The L3+Cosmics [36], which uses the L3 magnetic $\mu$ 
spectrometer and a scintillator setup to provide the $\mu$ arrival time,
has measured the muon spectrum in a wide energy range (20-2000 GeV) 
with a systematic error of 
2.6\% at $\sim 100$ GeV (mainly due to the uncertainty on the knowledge
of the 30 m thick molasse overburden) 
increasing to 15\% at 2 TeV due to the momentum resolution.
Using ALEPH hadronic calorimeter and TPC, the Cosmo-ALEPH experiment [37]
presented the muon flux and charged ratio
measurement in the range 70-2500 GeV. The results are preliminary since
the flux is normalized to the  world average and the experiment is 
investigating trigger efficiencies. 
Both results are of great interest for benchmarking atmospheric shower 
development codes (see Sec.~\ref{sec:calc}) up to high
energies. The L3+C measurement has already been compared to the
Bartol calculation in [16]. Other comparisons between calculations 
and balloon flight ascent data at $\lesssim 20$ GeV have been shown in 
[41,42,21].
The interest of data taken at the top atmosphere
(see for instance [41]) relies on the possibility of testing the
first interactions, when the shower has not yet developed.
\begin{figure}[t]
  \begin{center}
\begin{tabular}{cc}
    \includegraphics[height=19.pc,width=17.pc]{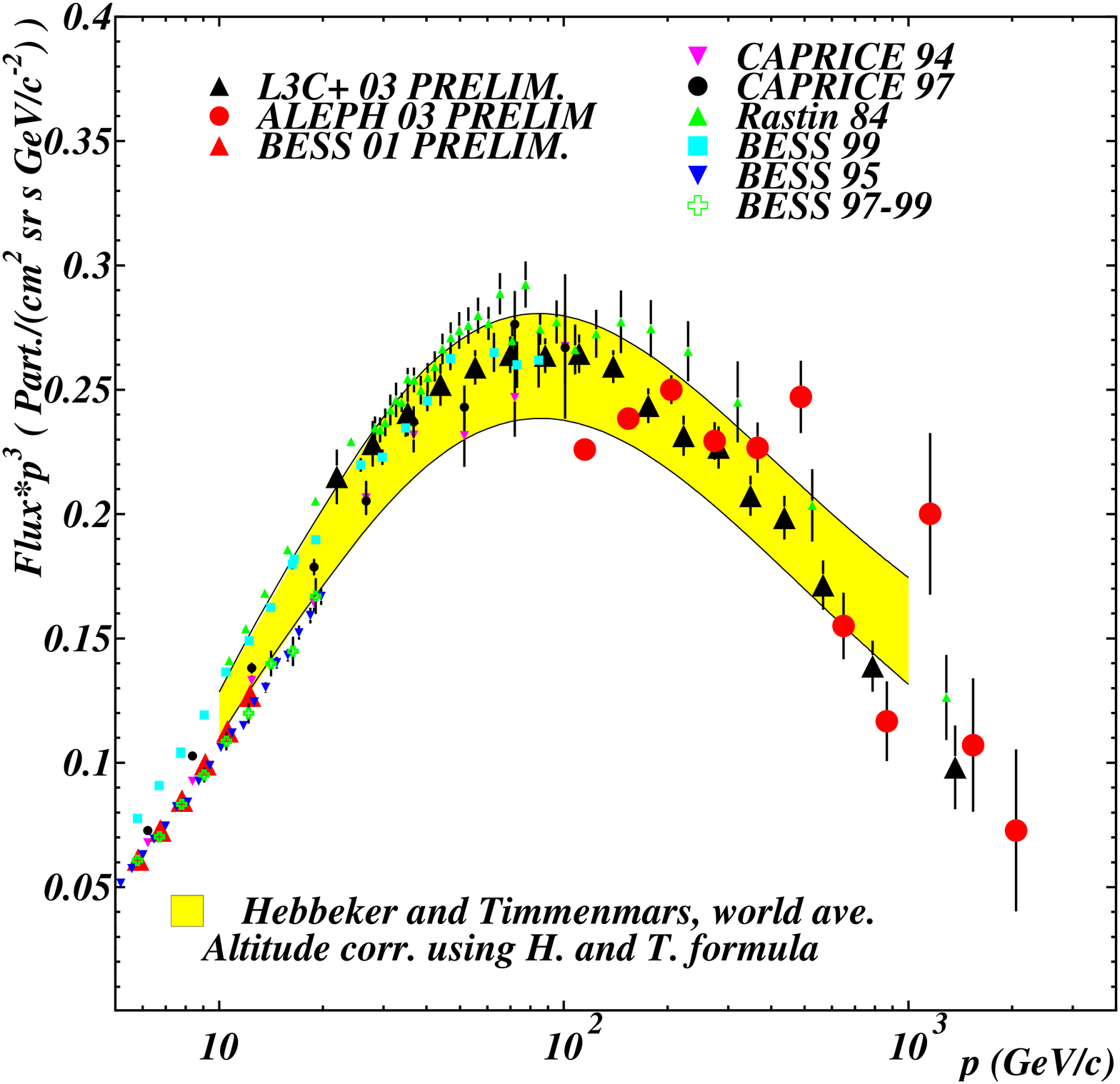} &
    \includegraphics[height=18.pc,width=16.5pc]{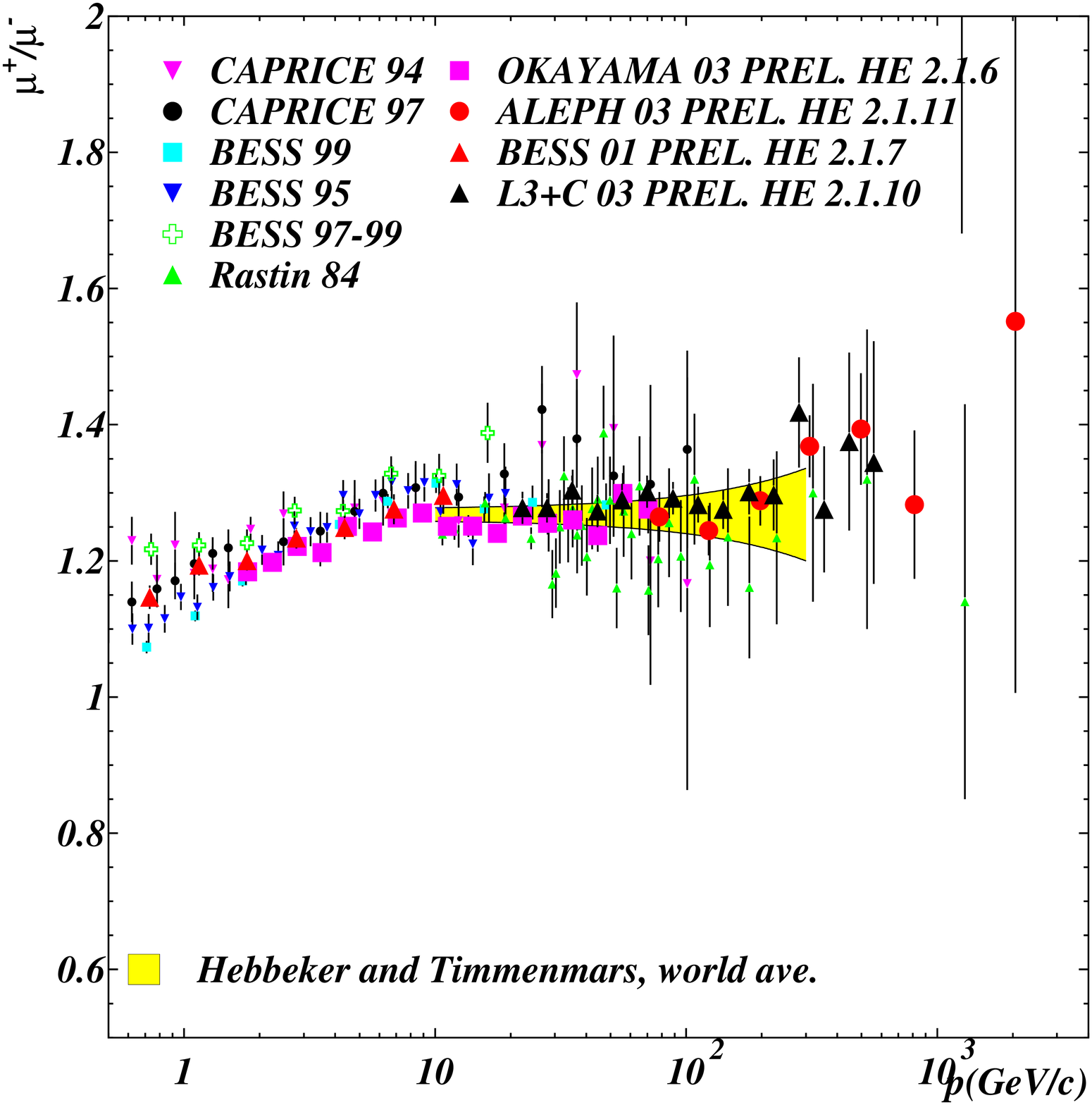} 
\end{tabular}
  \end{center}
  \vspace{-1.5pc}
  \caption{{\bf On the left:} Vertical atmospheric muon flux at sea level
(altitude corrected using the formula in [33]) 
vs momentum. The yellow band is the flux world average
in [33]. Data: L3+Cosmics (black triangles) [36], 
red full circles (Cosmo-ALEPH [37]), Rastin 84 (ref. in [33], green
triangles with the highest normalization at 100 GeV). At energies $< 10^2$ GeV:
BESS-01 at Ft. Sumner [35] (small red triangles), BESS 95 at Tsukuba 
(small blue reverse triangles) and BESS 97-99 at Lynn Lake (green crosses)
[38], BESS-99 at mountain altitude at Mt. Norikura (cyan squares) [39], 
CAPRICE 94 (pink reverse triangles not distinguishable from BESS 95) 
and CAPRICE 97 (black small circles) [34].
{\bf On the right:} Muon charge ratio: same symbols as the plot on the left, 
including also OKAYAMA (large pink squares) [40].}
\end{figure}

\section{Neutrino of Astrophysical Origin}
\label{sec:astro}

In this section results concerning measurements and searches for
neutrinos not produced in
the Earth atmosphere, such as solar neutrinos and neutrinos from
cosmic sources, are summarized. The only exception is for the reactor
experiment KamLAND,
that is included in the solar $\nu$ session due to the
impact of its results on the solar $\nu$ problem. 

\subsection{Solar Neutrinos}
\label{sec:solar}

In writing this section on solar neutrinos
it cannot be ignored the fact that after the conference 
the SNO collaboration has published results on the salt phase [43] 
and numerous interpretations have already appeared.
SNO, a 12 m diameter acrylic vessel filled with 1 kton pure $D_{2}$O  
seen by 9500 PMTs, exploits the solar $\nu$ detection above a $\sim 5$ MeV
threshold through 3 reactions: neutrino elastic scattering (ES),
mainly sensitive to $\nu_e$ since $\sigma_{\nu_e}
\sim 6 \sigma_{\nu_{\tau,\nu}}$, 
$\nu_e$ CC and all flavor $\nu$ NC interactions on deuterium. 
The ratio of CC to NC interactions provides the electron survival probability.
The results of the first phase of the experiment, also combined
with the SK precise measurement of ES showed that about
2/3 of $\nu_e$ convert into another active flavor $\nu_\mu$ or $\nu_\tau$,
indicating hence a clear preference for MSW effects [44] and a $5.3 \sigma$
rejection of the null hypothesis on flavor transformation.
After increasing sensitivity to NC with the addition of $\sim 2$ tons of salt,
the experiment is now able to claim evidence for MSW effects in solar 
neutrinos.
These results represent a fundamental step forward in understanding 
solar $\nu$ oscillations through the precise determination of
the total active $^{8}$B (and hep) flux in agreement with [45].  

SK [46] presented the updated analysis of 1496 days data taking 
with $E_{th}=5$ MeV. In this period
22385 solar $\nu$s were detected through the recoiling electron 
in ES interactions with sensitivity to $\nu$ energy and direction.
The day-night asymmetry parameter is $\frac{N-D}{(N+D)/2} 
= -0.021 \pm 0.020 \pm^{0.013}_{0.012}$, which is
consistent with zero, and the electron energy spectrum
is consistent with being flat with 44\% c.l.. No significant time variation
or energy distortion appears in the data favoring LMA (large mixing
angle) solution. Fig.\,4 (on the left) shows the allowed region at 95\% c.l.
including SK information on rate, spectrum and time-variations, SNO 
measurements of CC and NC available at the time of the conference [44], 
radiochemical experiments and the region allowed by the measurements of the 
rate and of the spectrum of reactor $\nu$s in KamLAND 
[47]. About 80\% of the muon neutrino flux reaches the
KamLAND detector (1 kton scintillator with a 34\% PMT coverage) 
from reactors at 140-210 km of distance with $<E_{\nu}> \sim 3$ MeV.
This disappearance experiment observed a deficit of events
with respect to the no-oscillation expectation 
of $\frac{N_{obs}-N_{backg}}{N_{exp}}
= 0.611 \pm 0.085_{stat} \pm 0.041_{sys}$ 
and measured the energy spectrum of prompt positrons from the
reaction $\bar{\nu}_e + p \rightarrow e^{+} + n$. 
The LMA region is divided into two allowed parts
including KamLAND results (if CPT is conserved).
After the results of SNO salt phase [43] the higher $\Delta m^2$ region 
disappears at 99\% c.l. and the best fit point is $\Delta m^2 =
7.1 \cdot 10^{-5}$ eV$^2$ and $tan^2\theta = 0.41$.
KamLAND [47] unfortunately has not presented new
results since most of the Japanese reactors have been running at
much reduced efficiency due to safety controls.
Preliminary results on anti-neutrinos produced in the Earth crust in U/Th
decays (geo-neutrinos with $E_{\nu}$ between 0.9-2.5 MeV) 
were presented posing a limit on the Earth heat source of 
$< 110$ TW (95\% c.l.), strongly dependent on the element 
distribution particularly of the Japanese Island Arc. 

A search for possible time modulations has been presented by SK after 
some papers appeared claiming for periodicities very close to
the time interval in which published data are binned (10 days) [48]. 
No significant periodicity (except for the long term modulation due to the
eccentricity of the Earth orbit around the Sun) 
is found and modulations of 10-100 (30-100) days
with amplitude larger than 10\% are excluded at 95\% c.l. binning data
in 5 (10) days.

Even though the LMA solution seems well established, the current
precision on parameters still allows for sub-dominant processes.
Limits on solar $\bar{\nu}_e$ appearance due to the
combined effect of spin flavor precession in presence of a 
large enough solar magnetic field and neutrino magnetic moment 
\footnote{Current best limit on the
$\nu$ magnetic moment is from the MUNU experiment $\mu_{\nu} < 10^{-10} \mu_B$ 
(90\% c.l.) [49].} (evolving $\nu_e$ into 
$\bar{\nu}_{\mu} \, , \bar{\nu}_{\tau}$) and of
flavor oscillations (converting 
$\bar{\nu}_{\mu} \, , \bar{\nu}_{\tau}\rightarrow
\bar{\nu}_e$) were presented by SNO [50] and SK [51].
This search addresses the fundamental question on the nature of
neutrinos, Majorana or Dirac particles. As a matter of fact, $\bar{\nu}_e$s
would be detected only if $\nu$s are Majorana particles, since
if they are Dirac ones the electron left-handed neutrinos would convert into
sterile (hence not detectable) right-handed ones.
The reaction exploited by SK is inverse $\beta$ decay, 
where positrons cannot
be separated by electrons and the coincident 2.2 MeV photons 
from neutron capture cannot be detected since their energy is below threshold
(contrarily to what happens in KamLAND [49]).
A 93\% contamination is expected
from spallation backgrounds [51]. On the other hand cleaner signatures
are found in SNO [50] from the reaction $\bar{\nu}_e + d \rightarrow 
e^{+} + n + n$ thanks to 3-fold and 2 fold (e$^{+}$n, nn) coincidences. 
Nevertheless, the efficiencies are only $\sim 1\%$ and 10\% respectively. 
In Tab.\,2 results are summarized including the post-conference
KamLAND result [49]. 
Furthermore, it should be considered that alternative $\bar{\nu}_e$ sources 
could be WIMP annihilation in the Sun, relic 
SN neutrinos ([52], see Sec.~\ref{sec:collapse}) and neutrino decay.
\begin{table}[t]
 \caption{Upper limits (90\% c.l.) on the flux of $\bar{\nu}_e$ in percentages 
of the Standard Solar Model $\nu_e$ $^8$B flux [45]. Live days and positron 
energy ranges are indicated.}
\begin{center}
\begin{tabular}{ccc}
\hline
SNO [50] & SK [51] & KamLAND [49]  \\  
\hline
1.02 & 0.8 & 0.028\\ 
($E_{th}=5$ MeV, 306 d)&
($E_{min}=8$ MeV, 1496 d)& ($E \in [8.3,14.8]$ MeV, 185.5 d) \\
\hline
\end{tabular} 
\end{center}
\end{table}

\subsection{Neutrinos from Stellar Collapse}
\label{sec:collapse}

About 99\% of the binding energy ($\sim 3 \cdot 10^{53}$ erg) of a collapsing 
star goes into neutrinos: $\nu_e$ are produced during the neutronization phase
which lasts about 10 ms, and neutrinos of all flavors 
during the thermalization phase of $\sim 10$ s.
Equipartition of energy luminosity between different $\nu$ flavors
and Fermi-Dirac spectra are expected, and average energies of 
$<E_{\nu_{e},\bar{\nu}_{e},\nu_{x}}> \sim 13,16,23$ MeV, even though recently
detailed simulations show that $<E_{\nu_e}>$ is much closer than 
before to $<E_{\nu_{\mu,\tau}}>$ [53]. 
The largest rate of events is foreseen for inverse $\beta$ reactions in
water/ice and scintillator detectors.
In Tab.\,3 the limits on the SN rates presented at
this conference and others are given.  As a reference, the
expected SN rate in our Galaxy is 2-4 per century. 

LVD [54] has presented an analysis on the effect on detected event rates due
to $\nu$ oscillations for the normal ($\Delta m^{2}_{32}>0$) and inverted 
hierarchy cases ($\Delta m^{2}_{32}<0$) which could lead in conservative 
models to an enhancement
up to $\sim 450$ events for a SN at 10 kpc 
emitting $2.5 \cdot 10^{53}$ erg in $\nu$'s,
or a suppression of events in pessimistic cases in which
$<E_{\nu_e}>$ is very close to $<E_{\nu_{\mu} , \nu_{\tau}}>$.
Essentially the enhancement is due to the fact that if $\bar{\nu}_{\mu , \tau}$
oscillate into $\bar{\nu}_e$, since they have
larger average energies, their spectra are harder than for $\bar{\nu}_e$
in the case of no oscillations.

\begin{table}[t]
 \caption{Limits (90\% c.l.) on the SN rate in units SN/yr 
for various experiments. Distances to which limits
apply and data taking times are indicated.}
\begin{center}
\begin{tabular}{cccc}
\hline
LVD [54] & SK [55] & AMANDA II [56]& MACRO [57]  \\  
\hline
0.24& 0.49  & 4.3 &0.27 \\
 $<20$ kpc, 10 yrs & $<100$ kpc, 4.7 yrs & in Galaxy, 80 hrs & 
in Galaxy, 10 yrs \\
\hline
\end{tabular} 
\end{center}
\end{table}

\subsection{The Neutrino Telescope Era}
\label{sec:nutelescopes}

The Neutrino Telescopes session (HE 2.3) contained numerous contributions 
indicating an intense experimental activity,
aiming at the detection of the first astrophysical $\nu$'s with 
high energies. These neutrinos 
would constitute new messengers from the universe,
since compared to photons which currently provide our best knowledge, 
they are less absorbed due to their weak interactions.
Pioneer works on neutrino astronomy (see for instance the review in [58])
were initiated by tracking calorimeters (Kolar Gold Field) and water Cherenkov
detectors (IMB, Kamiokande), followed by MACRO [59] and SK [60], with
areas of the order of $10^{2}-10^{3}$ m$^2$ and $E_{th}\sim 1$ GeV
looking for upward through-going muons produced by $\nu_{\mu}$ interactions
in the surrounding rock.

This detection technique profits of the increase with energy of the
$\mu$ range and the $\nu$ CC interaction. 
Typical astrophysical $\nu$ sources are beam dumps where protons are 
accelerated on matter or gas of photons producing charged mesons,
or decays of very massive particles, such as topological defects.
At high energies it is expected that the signal of $\nu$'s produced
in beam dumps with typical power law spectra from $1^{st}$ order Fermi 
acceleration mechanisms $E^{-2}-E^{-2.5}$, overcomes the steeper spectrum 
of the atmospheric $\nu$ background ($\sim E^{-3.6}$ above 100 GeV).
The rejection of the atmospheric $\mu$ background is achieved by
measuring the up-going $\mu$ direction and by locating
detectors below kilometers of matter.
Pointing capabilities and energy measurement are relevant to
reject both sources of background.

The low expected event rates from astrophysical sources 
and current experimental upper bounds, urged
the construction of huge neutrino telescopes 
(areas of the order of 0.01-1 km$^2$,   
located in the South Pole ice/lake/sea water depths). Essentially these
are 3-dimensional arrays of PMTs that allow to reconstruct $\mu$ tracks
in water using the times of PMTs hit by the emitted Cherenkov photons 
and to estimate the energy from charge measurements.
Besides cascades from NC, CC $\nu_e$ and $\nu_{\tau}$ interactions
(about 1/2 of $\nu_{\mu}$ from cosmological sources oscillate into
$\nu_{\tau}$ assuming atmospheric $\nu$ oscillation parameters)
are detected as point-like sources of light.
The extreme environmental conditions of the locations of the
experiments represent a challenge and the field is 
rich of successes and drawbacks. At this conference numerous 
results have been 
presented by AMANDA [61] and Lake Baikal [62]. Supported by the 
experience acquired with AMANDA, the challenge at the South Pole
will continue with an improvement of more than 2 orders of magnitude
in sensitivity with the IceCube detector, whose construction will
start in the austral summer 2004-5 and will last 6 yrs [63]. 
The community is now looking forward to see first results from experiments
in the Mediterranean: ANTARES [64] will be completed in 2006 and common efforts
between ANTARES, NESTOR [65] and NEMO-RD [66] could lead to the construction
of a km$^3$ detector which would complement IceCube sky coverage.
The need for huge detectors is also motivating R\&D on new techniques 
alternative
to the Cherenkov one, such is the case of the RICE radio detector [67].
\begin{figure}[t]
  \begin{center}
\begin{tabular}{cc}
    \includegraphics[height=18.pc,width=17.pc]{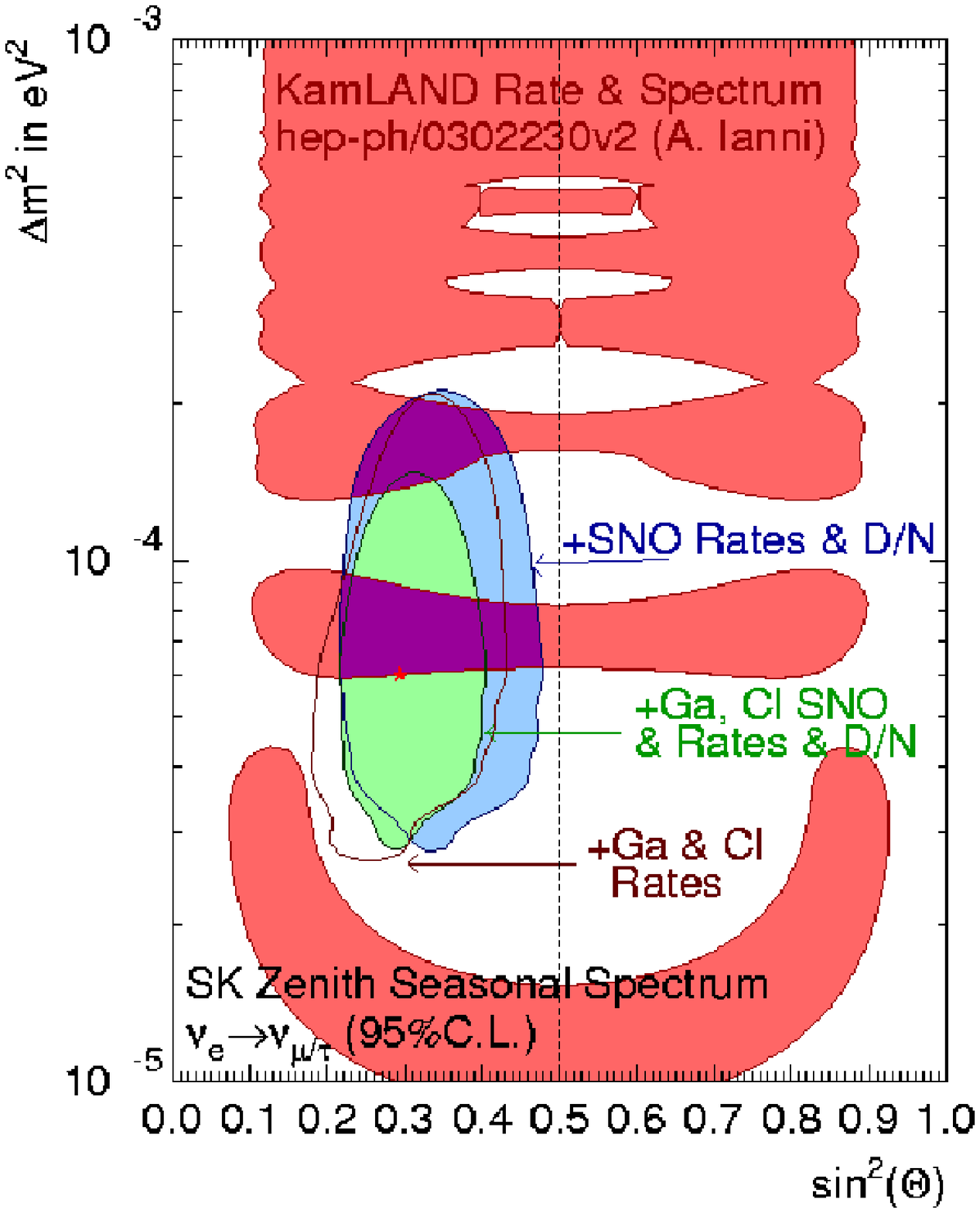} &
    \includegraphics[height=18.pc,width=17.pc]{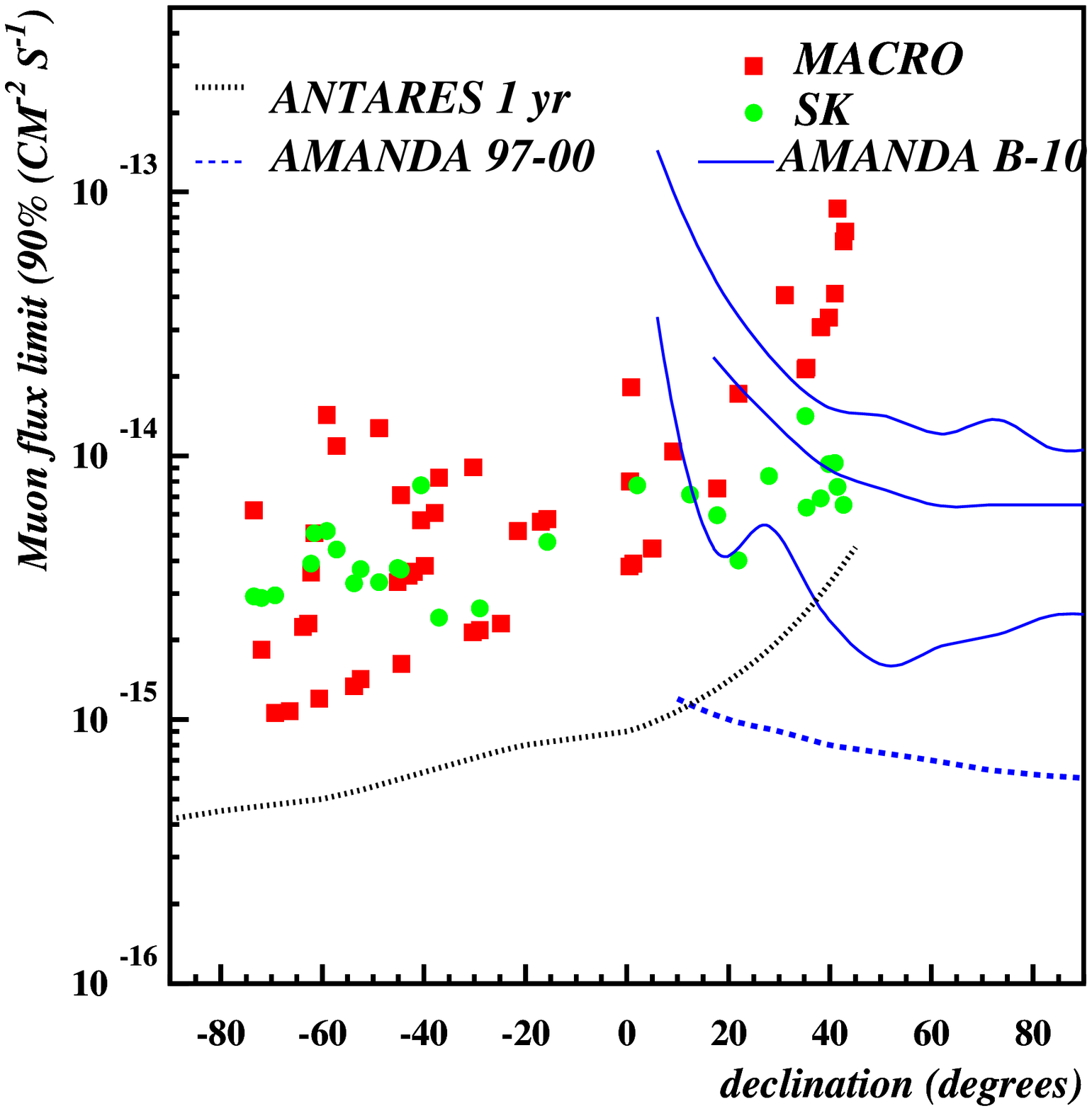} 
\end{tabular}
  \end{center}
  \vspace{-1.5pc}
  \caption{{\bf On the left:} Allowed region (95\% c.l.) from the combination of SK
[46], SNO [44], Ga and Cl experiments. Superimposed the darker
allowed region from KamLAND [47].
{\bf On the right:} 90\% c.l. upper limits on the muon flux induced 
by neutrinos with spectrum of the form $E^{-2}$ as a function of source
declination for SK (green 
circles)[60], MACRO (red squares) [59], AMANDA-B10 (3 blue solid lines:
the upper and lower line give the maximum variations of limits for
different right ascension bins in the same declination band) [69], AMANDA-II
(dashed line) foreseen limit for 1997-2000 data sample [68],
ANTARES sensitivity (dotted black line) after 1 yr [76].
It has not been possible to apply a correction due to different
muon average energy thresholds in this plot (SK $\sim 3$ GeV, MACRO
$\sim 1.5$ GeV, AMANDA $\sim 50$ GeV). Nevertheless, the 
maximum of the response curves for all of these
detectors for an $E^{-2}$ flux is $E_{\mu} \sim 1-10$ TeV, hence the
number of events contributing between 1-50 GeV should not
make a large correction to these limits. Moreover, $\mu$ flux limits
are less dependent on different $\nu$ flux models than $\nu$ limits.}
\end{figure}
\subsection{Status of the Experiments and Results}

AMANDA has presented first results of the current AMANDA-II
configuration made of 677 optical modules (OM), transparent pressure resistant
spheres containing PMTs, distributed over 
19 strings inside a cylinder of 120 m diameter located at depths between 
1500 and 2000 m in the polar ice. 
This layout increased the sensitivity in the
high energy range of about a factor of 4 compared to AMANDA-B10
configuration active during 1995-7, with 302 OMs on 10 strings.
The effective area for $\nu$-induced muons is now almost uniform with 
declination being on average 
$\sim 0.03$ km$^2$ at 10 TeV. The angular resolution 
benefits from the additional strings,
being now about $2.3^{\circ}$ [68] compared to about $4^{\circ}$ [69] for 
AMANDA-B10.
The absolute pointing precision is at the level of $0.5^{\circ}$ verified using
the SPASE extensive air shower array (EAS) at the surface.

Two test-beams can be used by these experiments to 
understand systematics on efficiencies and acceptance:
atmospheric muons and neutrinos.
The shape of atmospheric $\mu$ vertical intensity as a function of the 
cosine of the zenith of 10 hrs data is in agreement with 
the simulation using CORSIKA with the QGSJET interaction model, but the
data need to be reduced by 30\% to agree on the normalization [70]. This is
compatible with the systematics of the experiment mainly due to
depth-dependent ice optical properties and bubbles formed around OMs 
after drilling and hence to OM sensitivities.
Also the energy spectra of both beams have been reconstructed
with a muon energy resolution (defined as the standard deviation 
of the $log_{10}E$ distribution) of 0.4-0.6 between 500 GeV-5 PeV.
In particular in the 1 TeV region, AMANDA atmospheric 
$\nu$ spectrum, having a fitted
slope of $-3.56 \pm 0.20_{stat}$, agrees with the Fr\'ejus one [71]. 
Also a method has been developed 
to extrapolate the CR spectrum in the 1.5-200 TeV/nucleon 
region, independently of ice optical properties, OM sensitivities and detector
configuration changes. This method results in a spectral
index and a normalization for protons of $-2.80 \pm 0.02$ and 
$0.106 \pm 0.007$ m$^{-2}$ s$^{-1}$ sr$^{-1}$ TeV$^{-1}$, respectively [72]. 
The result is in reasonable agreement with direct measurements.

AMANDA experience is precious for the construction of
the km$^3$ scale detector, IceCube [63], with 4800 DOM (Digital Optical 
Modules) on 80 strings spaced by $\sim 125$ m, implemented between 
1400-2400 m below the surface.
DOMs will exploit digital readout of 10 inch PMTs storing the full waveform 
with a dynamic range of 200 photoelectrons per 15 ns.
The detector will be complemented by
the EAS array IceTop, 80 stations close to IceCube holes made 
of 2 tanks filled with ice seen by DOMs [73]. 
IceCube declared effective area after selection cuts reaches 1 km$^2$ 
at 10 TeV and the angular resolution is about $0.6^{\circ}$
above 100 TeV, improving with energy. For energies $\gtrsim 1$ PeV 
upgoing $\nu_{\mu}$ are non negligibly absorbed in the Earth depending
on the zenith angle. Thanks to energy cuts getting rid of
atmospheric $\nu$s and $\mu$s, the experiment will be able to
measure astrophysical down-going $\nu$ events.

Results from the radio-frequency detector RICE [67] were quite impressive. 
The detector is made of 16 radio receivers with a frequency bandwidth 
$\sim 200-500$ MHz located in holes drilled
for AMANDA at depths between 100-300 m  
over a volume of about (200 m)$^3$.
The attenuation length in ice for radiation is $>1$ km. 
The technique exploits the
detection of few ns radio pulses produced by $\nu$ induced electro-magnetic
cascades. A negative charge excess develops in the shower 
due to $e^+$ annihilation
and extraction of $e^-$ from the media resulting in a 
coherent Cherenkov emission (Askarian effect) 
proportional to the square of the primary particle energy. 
Limits comparable to AMANDA-II sensitivity were obtained using such
a cost effective technique. Given that a better rejection of
the anthropogenic noise could be achieved deploying receivers deeper
in the ice, the opportunity to implement
receivers also in IceCube holes should be considered.

In the Northern hemisphere, 
the NT-200 Baikal detector is running at 1100 m depth in the Siberian 
Lake Baikal [62], with about 192 OMs on 8 strings. The angular resolution
is about $3^{\circ}$. In Mar. 2003 a prototype string has been deployed
at about 100 m from the center of the detector with 6 couples of OMs; 
with 2 more strings the sensitivity to
cascade events will be improved by a factor of 4.
Baikal has measured 84 neutrino events (268 d for NT-200 and
70 d for the previous configuration NT-96) compatible with the hypothesis
of being of atmospheric origin.
   
ANTARES [64], a European project started in 1996,
is deploying a 12 string detector with a total of 900 OMs,
containing 10 inch PMTs, in the Mediterranean close to the South France
coasts. The effective area, estimated after selection
requirements on the track reconstruction error, is $> 0.02$ km$^2$
for $E_{\nu} > 10$ TeV 
and the angular resolution achieves a limiting value of 
$\sim 0.2^{\circ}$ for $E_{\nu} > 10$ TeV, mainly due to 
the transit time spread of PMTs and the diffusion of light in water.
The 'junction box', transmitting power and
data to strings, is already lying on the sea bed at 2500 m below the surface
since Dec. 2002.
It is connected to a 40 km long electro-optical cable deployed in Oct. 2001.
Two strings have been deployed by a manned submarine 
and operated between Dec. 2002 and Jul. 2003: 
one prototype string including 15 OMs with the final front-end electronics and
the other string for environmental parameter measurements [74].
The prototyping experience allowed to verify the detector
design and functionality, to find solutions to a few occurred problems,
and yielded a vast amount of measurements on 
$^{40}K$ and on a strongly variable 
bioluminescence activity. It was found that this environmental background
rate is below 200 KHz for $>90\%$ of the time.
 
In Mar. 2003 NESTOR [65] deployed 10 km off of Pylos coast (Greece)  
a 12 PMT test floor of reduced dimensions
compared to the hexagonal detector design (6 couples of
up-down looking PMTs are located at 6 m from the center) at $\sim 4000$ m 
depth connected to shore by a 28 km cable. Cable connection operations 
performed in air
on a boat and data readout and transmission were successful. 

Fig. \,4 (on the right) summarizes current upper limits on $\nu$-induced 
$\mu$ fluxes for point-like sources and expected sensitivities
as a function of their declination.
The AMANDA-II sample [68] for point-like source searches consists in 
699 upgoing events. Upper limits were given for a few sources and are
at the level of a few $10^{-15}$ cm$^{-2}$ s$^{-1}$. 
The estimated level of limits for the 97-02 data sample
is shown together with published results from AMANDA-B10 [69].
Also shown are the limits by SK [60] using 2369 
through-going and stopping muons with an angular resolution of about 
$4^{\circ}$.
The 354 showering events, used for Weakly Interacting Massive
Particles (WIMPs) in [75], with estimated average energy of 
$\sim 1$ TeV could represent a 'gold' sample for astrophysics searches.
The expected sensitivity after 1 yr of data taking of ANTARES
is given in [76], where a comparison of binned and unbinned
likelihood ratio methods is performed.

In Fig.\,5 (on the left) 
results on upper limits on $\nu_{\mu}$ diffuse fluxes  
compared to models are shown. The atmospheric backgrounds are
rejected through energy cuts and $\mu$ track reconstruction
is required except for the Ultra-High Energy (UHE) AMANDA analysis [77].
The absorption of $\nu_{\mu}$ in the Earth is included.

Cascades produced by $\nu_e$, $\nu_{\tau}$ CC and NC interactions 
are detected with worse angular resolution but with better energy
reconstruction than $\nu_{\mu}$ events
(e.g. for AMANDA $30^{\circ}-40^{\circ}$ and log E resolution of
0.1-0.2 between $\sim 50$ TeV $-100$ PeV).
Cascade events are detected from the entire solid angle, 
the background discrimination being possible thanks to vertex identification.
Limits on diffuse fluxes of all flavor neutrinos are shown in Fig. \,5 
(on the right).

Many estimates on $\nu_{\tau}$ astrophysical fluxes produced after
oscillations, $\nu_{\tau}$ regeneration processes and $\tau$ propagation 
in the Earth have been presented up to the EeV energy range [81,82,83,84,85]. 
Even though $\nu_{\tau}$s are not absorbed in the Earth due to 
their regeneration chain ($\nu_{\tau}$ CC 
interaction followed by fast $\tau$ decay which produces back $\nu_{\tau}$
and other flavor $\nu$'s), they loose energy. Hence
event rates are low since $\tau$ leptons can be recognized 
at energies $>1$ PeV, when $\tau$ range in water/ice is $> 50$m.
The $\tau$ events can be unequivocally identified if the 2 cascades 
from CC interaction and decay (double bang events) 
are detected but only a few events/yr/km$^2$ are expected [85].
Most of the UHE events ($>0.1$ EeV) can be detected
from the upper hemisphere in an IceCube-like detector.
For a GZK $\nu$ model, in which 
UHECR make photopion interactions on the cosmic microwave 
background, $\sim 50$ $\mu$ and $\tau$ events in a km$^{2}$ per year 
are expected from the upper hemisphere and a few from the lower
[81].

The conversion efficiency of $\nu_{\tau}$ into $\tau$
for $\sim 10$ km of rock is much larger above 1 PeV than 
in $\sim 3000$ g/cm$^2$ of horizontal atmosphere [84]. Hence
event rates in fluorescence and Cherenkov arrays 
due to Earth skimming $\nu$s ($\nu_{\tau}$s interacting in
Earth chords of the order of a few tens of kms) or mountain
events ($\nu_{\tau}$s crossing a few tens of kms of mountain rocks
and producing a shower detectable from another peak $\sim 10$ km 
far from the mountain) are more encouraging than event rates produced by
horizontal $\nu$s interacting deep in the atmosphere [84,86].
\begin{figure}[t]
  \begin{center}
\begin{tabular}{cc}
    \includegraphics[height=19.pc,width=17.pc]{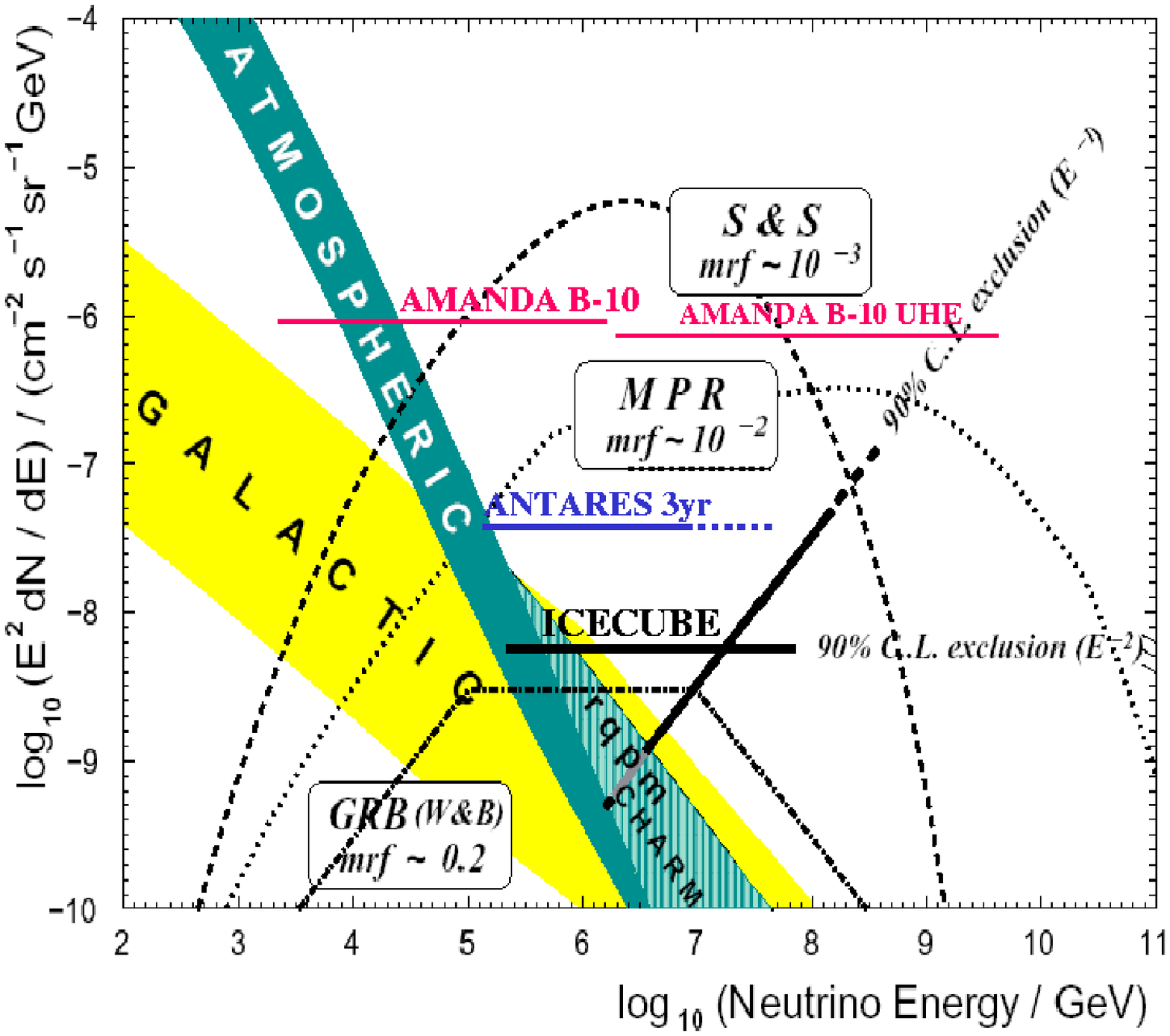} &
    \includegraphics[height=19.pc,width=17.pc]{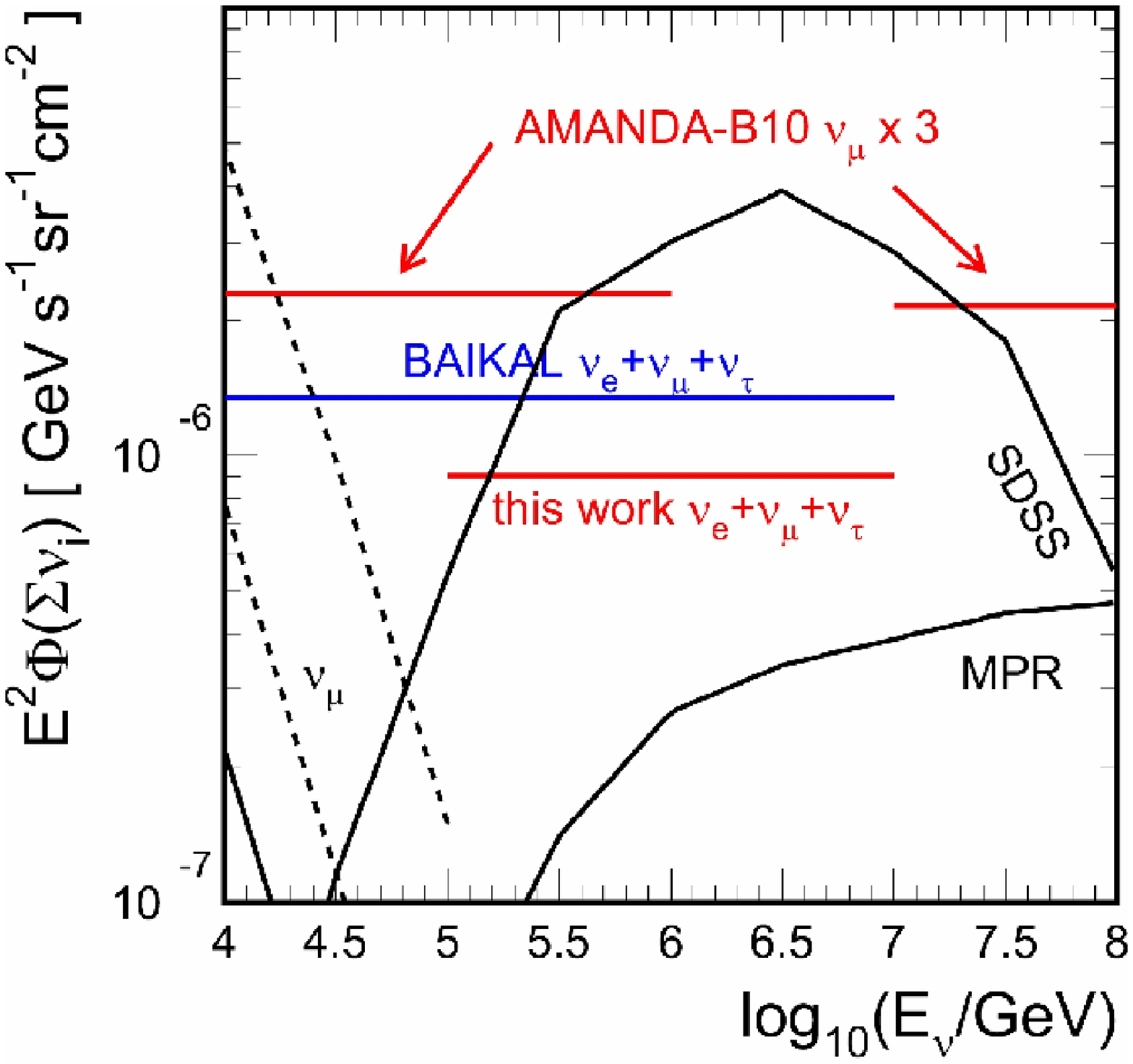} 
\end{tabular}
  \end{center}
  \vspace{-1.5pc}
  \caption{{\bf On the left:} 90\% c.l. upper limits on the $\nu_{\mu}$ diffuse flux
vs $E_{\nu}$. Horizontal lines in this units are limits for
$E^{-2}$ $\nu$ fluxes. Experimental limits
are indicated by the name of the experiment: 
AMANDA B-10 (130 days) limit up to 1 PeV [78]
and the limit for UHE horizontal showering $\mu$s up to a few EeV [77];
sensitivities for 3 yrs of data taking of ANTARES [79]
and of IceCube [64] 
(also shown for an $E^{-3}$ flux).  
Also some values of the
mrf (the model rejection factor is 
defined as the average upper limit at 90\% c.l. for many
possible experimental outcomes over the expected signal)
for IceCube are indicated. Notice that
even if the Waxman and Bahcall prediction for GRBs [100] appears to be below 
IceCube sensitivity, the search for $\nu$ signals in coincidence
with accompanying GRBs is background free, resulting in a mrf $<1$.
{\bf On the right:} 90\% c.l. upper limits on diffuse $E^{-2}$ $\nu$ flux
for cascades of all flavors. The AMANDA-II (197 d, indicated by 'this work') 
[80] and Baikal (268 d)
[62] results are presented compared to atmospheric $\nu$ fluxes,
an AGN model (SDSS) 
and an upper limit extrapolated from CR UHE measurements (MPR). 
For comparison the AMANDA B-10 $\nu_{\mu}$
limits are shown multiplied by a factor of 3 (the underlying hypothesis
is that at source the proportion between flavors is $\nu_e : \nu_{\mu} :
\nu_{\tau} = 1:2:0$, while at Earth, after propagation and oscillations,
it is $1:1:1$).}
\end{figure}

\section{Dark Matter Searches}
\label{sec:dm}

After the publication of WMAP data [87] cosmology entered a new precision era. 
About 26.7\% of the universe density is due to matter, most of
which is dark and only 4.4\% is of baryonic nature; 
the remaining 73.3\% is due to dark energy.
Most of the searches for dark matter (DM) presented at the conference 
are indirect in the sense that secondaries produced by 
dark matter annihilation are looked for, except for the Bulby mine
experimental program on DM direct detection (DM particles
interact in the detector) described in [88].

The most interesting candidates for Cold DM are WIMPs, 
since their annihilation cross section
at the weak scale would account for the DM in the Universe. Between
CDM particles, the Supersymmetric (Susy) neutralino, a linear combination
of Susy partners of the photon, Z$^0$ and neutral Higgs bosons, 
is one of the favorite candidates.
Indirect searches for dark matter are performed by satellite, balloon
and ground-based Cherenkov detectors looking for $\bar{p}$, anti-deuterons,
$e^+$ and $\gamma$ excesses respect to the expected secondary fluxes 
produced by CR interactions during diffusion in the Galaxy.
Two intriguing indications of a DM component in secondary fluxes 
have stimulated a lot of discussions in the field.
EGRET has measured a diffuse gamma-ray flux in excess compared to 
standard models of primary CR interactions with the interstellar
medium indicating the possible presence of a diffuse source in the
Galactic Center region [89].
Moreover the HEAT balloon flights in 1994-95 measured a not highly
significant positron
excess above secondary production models around $\sim 10$ GeV [90].
The AMS-02 experiment, to be installed on the International Space Station
in 2005, will measure all these channels up to a few hundreds of GeV [91].  
As an example the performances for the $e^+$ channel
are shown in Fig.\,6 (on the left).
Anti-proton fluxes from neutralino annihilation
using spherical isothermal distributions of DM in the halo
indicate that it would be difficult to single out the DM contribution
in the secondary background and to constrain the Susy
parameter space [92]. Moreover the estimates are affected by uncertainties on
propagation models. More optimistic predictions could be obtained 
using different density profiles or hypothesis about clumpy halos.

A cleaner DM signature with respect to excesses in diffuse fluxes 
can be monoenergetic $\gamma$ lines produced by neutralino annihilation
($\chi \chi \rightarrow \gamma \gamma$,$\chi \chi \rightarrow \gamma Z$).  
Searches for this signal have been presented by HEGRA [93] and Milagro [94].
The HEGRA system of imaging atmospheric Cherenkov telescopes
surveyed the Andromeda galaxy (M31) looking for 500 GeV-10 TeV 
gamma-ray emissions
with an energy resolution of 10\%, and derived upper limits still far from
models not including dark matter clumpiness.
Milagro water Cherenkov detector has looked for TeV gamma-rays
from the Sun neighborhood possibly due to neutralino trapped in the solar
system.

Results on indirect DM searches presented by neutrino telescopes
[95, 75, 62, 102] are summarized in Fig.\,6 for $\nu$ induced $\mu$ fluxes 
produced by neutralino annihilation in the core of the Earth (on the right).
\begin{figure}[t]
  \begin{center}
\begin{tabular}{cc}
    \includegraphics[height=19.pc,width=16.pc]{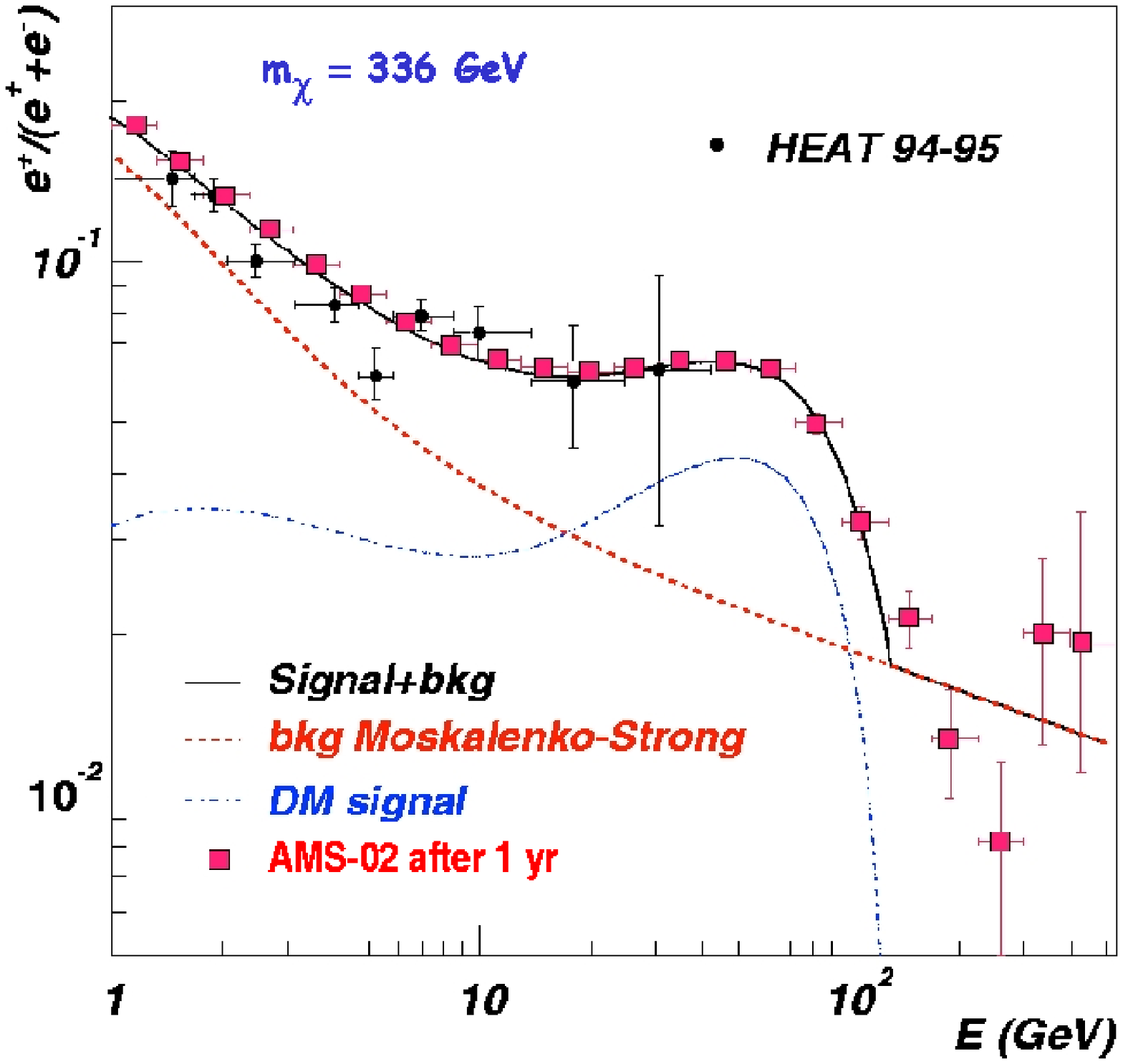} &
    \includegraphics[height=19.pc,width=18.pc]{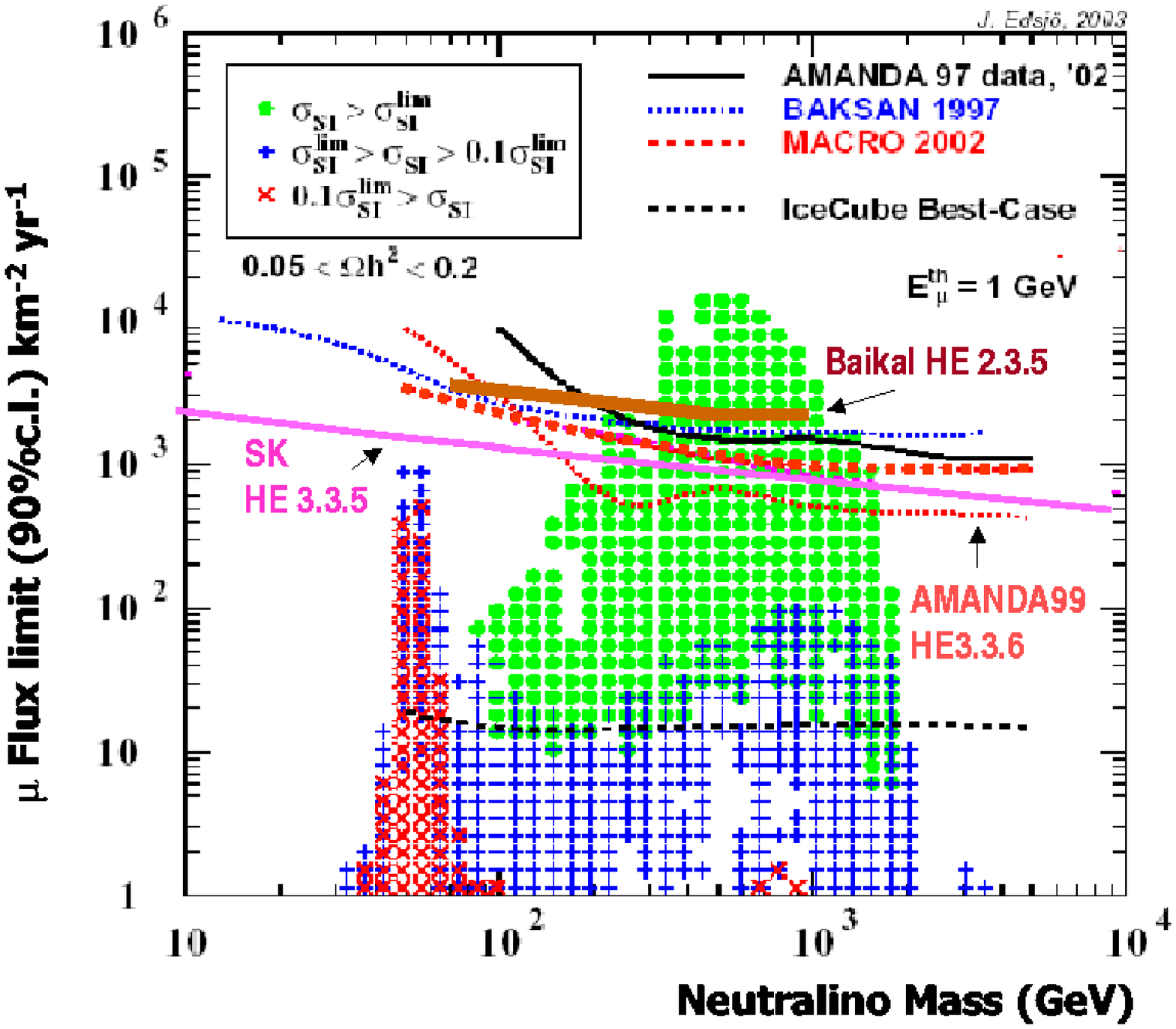} 
\end{tabular}
  \end{center}
  \vspace{-1.5pc}
  \caption{{\bf On the left:} fraction of $e^+$ measured by HEAT (circles), 
and AMS expected performances (squares) [91]. The solid line is the sum
of the expected secondary spectrum from standard propagation models (dashed
line) and the signal expected from annihilation of neutralinos
with mass of 336 GeV. The signal has been
multiplied by a factor of 11.7 to fit the
HEAT data [90]. {\bf On the right:} $\nu$ induced $\mu$ fluxes 
induced by neutralino annihilation in the core of the Earth. Susy models
are calculated using DarkSUSY [96] and the different symbols used
indicate: models excluded by Eidelweiss direct search experiment [101]
(green circles), models that would be excluded with a factor of 10 increased 
sensitivity (blue crosses) and those which would require a larger
sensitivity.
Experimental limits are scaled to an energy threshold of 1 GeV.
The limits presented at the conference are indicated with the
name of the experiment: AMANDA B-10 [95], SK [75] and Lake Baikal [62].
The dashed red line is the MACRO limit [97].}
\end{figure}
\section{Conclusions}

At this conference, many refinements on experimental analysis
have been presented in the neutrino sector, even though no striking news. 
Atmospheric $\nu$ experiments strongly indicate
that muon neutrinos oscillate maximally into tau neutrinos.
However, Super-Kamiokande oscillation parameter
best fit value has changed a little after the analysis update.
Naturally the question on how much the oscillation parameter estimate
is affected by the knowledge of atmospheric $\nu$ beam arises,
a subject widely discussed at the conference where many improved
atmospheric $\nu$ calculations were presented. Still uncertainties
are at the level of 15-30\% on absolute fluxes, not affecting, however,
the robustness of the result in favor of oscillations.
Solar $\nu$s are producing striking results in these years
and provide a useful mean to investigate fundamental properties
on $\nu$s, such as the question if it is a Dirac or Majorana particle.
This is one of the still open problems which should be addressed with
efforts comparable to those which are going to be devoted
on $\nu$ mixing matrix elements determination.
The Dark Matter problem also is such a fundamental question
that urge big efforts to find new evidences and to
confirm already existing indications, such as the DAMA result [98].

\section*{Acknowledgments}
I would like to thank the organizers of this conference, and in particular 
T. Kajita, for inviting and supporting me. The efficiency of the organization
was really impressive. 
An ICRC rapporteur needs friends that help, and for this I thank F. Cafagna. 
Also I thank I. Sokalski,
G. Battistoni, T.K. Gaisser, E.Lisi, F. Ronga, P. Lipari, M. Honda, A.
Habig, A. Marini, S. Cecchini,
F. Arneodo and all the people which explained me their works and 
taught me a lot.

\section{References}
\vskip -0.5 cm
{\small
\vspace{\baselineskip}
\re
1.\ Koshiba M.\ 2003, {\it Birth of Neutrino Astrophysics}, Hess lecture, 
this conference
\re
2.\ Habig A. for Super-Kamiokande Coll. 
\ 2003, {\it Atmospheric Neutrino Oscillations in SK-I},
HE 2.2.8
\re
3.\  Kasuga S.\ et al.\ 1996, Phys. Lett. {\bf B374}, 238
\re
4.\ See for instance: {\it A study of $\nu_{\mu} \leftrightarrow \nu_{\tau}$ vs.
$\nu_{\mu} \leftrightarrow \nu_{s}$ neutrino oscillation in 
atmospheric neutrinos
using a K2K near detector measurement}, C. Mauger, PhD Thesis, Nov. 2002,
available at http://www-sk.icrr.u-tokyo.ac.jp/doc/sk/pub/index.html    
\re 
5.\ Honda M.\ et al.\ 2003, {\it A Precise Three-Dimensional Calculation of the
Atmospheric Neutrino Flux}, HE 2.4.2 and 2001, Phys. Rev. {\bf D64} 053011
\re
6.\ Gaisser T. K., Honda M., Lipari P. and Stanev T.S.\ 2001, {\it Primary 
spectrum to 1 TeV and beyond}, Proceedings of 27$~{th}$ 
Int. Cosmic Ray Conf. (ICRC2001), 
Hamburg, 7-15 Aug. 2001, http://www.copernicus.org/icrc/papers/ici6694\_p.pdf
\re
7.\ Kajita T.\ 2003, {\it Super-Kamiokande Evidence for Muon-Neutrino 
Oscillations}, to appear in Proc. of X Int. Workshop on 
``Neutrino Telescopes'', Mar. 11-14, 2003, Venice
\re
8.\ Ambrosio M.\ et al.\ 2003, {\it Measurements of Atmospheric Muon Neutrino
Oscillations with MACRO. Conclusive analysis of the data collected with MACRO},
in preparation
\re 
9.\ Sanchez M.\ et al.\ 2003, {\it Observation of Atmospheric Neutrino Oscillations
in Soudan 2}, subm. to Phys. Rev. {\bf D} and hep-ex/0307069
\re
10.\ Sala P.R. for ICARUS Coll.\ 2003, {Status of the ICARUS 
Project}, HE 2.5.2
\re
11.\ Moriyama S. for Super-Kamiokande Coll. 2003, {\it Characterizing the
Atmospheric Neutrino Flux}, HE 2.2.10
\re
12.\ Lipari P.\ 2000, Astrop. Phys. {\bf 14}, 171 
\re
13.\ Nakayama S. for Super-Kamiokande Coll.\ 2003 
{\it Study of Atmospheric Neutrino Oscillations Using $\pi^0$ Events in SK-I},
HE 2.2.9
\re
14.\ Saji C. for Super-Kamiokande Coll.\ 2003 
{\it Search for Charged Current Tau Neutrino Appearance in Super-Kamiokande}, 
HE 2.2.11
\re
15.\ Gaisser T.K. and Honda M.\ 2002, Ann. Rev. Part. Sci. {\bf 52}, 153
\re 
16.\ Engel R.\ et al.\ 2003, {\it TARGET 2.2 - A Hadronic Interaction Model 
for Studying Inclusive Muon and Neutrino Fluxes}, HE 3.1.7
\re
17.\ G. Barr  \ et al.\ 2003, {\it A 3-dimensional Atmospheric Neutrino Flux
Calculation}, 1-P-272 
\re
18.\ Battistoni G., Ferrari A., Montaruli T. and Sala P.R. 2003, 
{\it High Energy Extension of the FLUKA Atmospheric Neutrino Flux}, 1-P-270,
all references and flux tables in 
http://www.mi.infn.it/\%7ebattist/neutrino.html;
FLUKA official WEB page: http://www.fluka.org
\re
19.\ Engel R.\ 2003, {\it Influence of Low-Energy Hadronic Interaction
Programs on Air Shower Simulations with CORSIKA}, HE 2.1.5
\re
20. Wentz J.\ et al.\ 2003 {\it Simulation of Atmospheric Neutrino Fluxes with
CORSIKA}, 1-P-271
\re
21.\ Derome L., Liu Y. and Buenerd M.\ 2003, {\it 3-Dimensional Simulation of
Atmospheric Muon and Neutrino Flux}, HE 2.4.1 
\re
22.\ Alcaraz J.\ et al.\ 2000 Phys. Lett. {\bf B490}, 27 and
2000 Phys. Lett. {\bf B494}, 193 
\re
23. Sanuki T.\ et al.\ 2000 Astrop. J. {\bf 545}, 1135
\re 
24.\ Boezio M.\ et al.\ 1999, Astrop. J. {\bf 518}, 457 
\re
25.\ Boezio M.\ et al.\ 2003, Astrop. Phys. {\bf 19}, 583
\re
26.\ Asakimori K.\ et al.\ 1998, Astrop. J. {\bf 502}, 278 
\re
27.\ Makoto H. for RUNJOB Coll., 
{\it Primary Proton and Helium Spectra Observed by RUNJOB}, OG 1.1.17
\re
28.\ Apanasenko \ et al.\ 2001, in Proc. of $26^{th}$ Int. Cosmic Ray
Conf. (ICRC99), Salt Lake City, 17-25 Aug. 1999
\re
29.\ Agrawal V., Gaisser T.K., Lipari P. and Stanev T.S.\ 1996, Phys. Rev
{\bf D53} 1314
\re
30.\ Zatsepin V.I.\ et al.\ 2003, {\it Rigidity Spectra of Protons 
and Helium as
Measured in the First Flight of the ATIC Experiment}, OG 1.1.15
\re
31.\ Battiston R., Rapporteur talk on OG1.1-2, OG1.5, this conference
\re
32.\ Bertaina M. for EAS-TOP and MACRO Coll.\ 2003,
{\it The Proton, Helium and CNO Fluxes at $E_{0} \sim 100$
 TeV from the EAS-TOP
(Cherenkov) and MACRO (TeV Muon) Data at the Gran Sasso Laboratory},
HE 1.1.15
\re
33.\ Hebbeker T. and Timmermans C.\ 2002, Astrop. Phys. {\bf 18}, 107
\re
34.\ Kremer J.\ et al.\ 1999, Phys. Rev. Lett. {\bf 83}, 4241 
\re
35.\ Tanizaki K. for the BESS Coll.\ 2003,
{\it Geomagnetic Cutoff Effect on Atmospheric Muon Spectra at Ground Level}, 
HE 2.1.7, Abe K.\ et al.\ 2003, Phys Lett. {\bf B564}, 8 
\re
36.\ Unger M. for the L3+Cosmic Coll.\ 2003, {\it Measurement of the 
Atmospheric Muon Spectrum from 20 to 2000 GeV}, HE 2.1.10
\re
37.\ Zimmerman D. for the CosmoALEPH Coll.\ 2003, {\it The Cosmic Ray Muon
Spectrum and Charge Ratio in CosmoALEPH}, HE 2.1.11
\re
38.\ Motoki M.\ et al.\ 2003, Astrop. Phys. {\bf 19}, 113
\re
39.\ Sanuki T.\ et al \ 2002, Phys. Lett. {\bf B541}, 234
\re  
40.\ Tsuji S. for OKAYAMA Coll.\ 2003, 
{\it Atmospheric Muon Measurement at Sea Level IV:
Muon Charge Ratio}, HE 2.1.6
\re
41.\ Abe K.\ et al.\ 2003, {\it Calculation of Muon Fluxes at the Small Atmospheric 
Depths}, HE 2.4.6 
\re
42.\ Stanev T.S.\ et al.\ 2003, 
{\it Comparison between CAPRICE98 Atmospheric 
Muon Data and Simulations with TARGET}, HE 2.4.9
\re
43.\ Ahmed S.N.\ et al.\ 2003, {\it Measurement of the total active $^{8}$B
Solar Neutrino Flux at the Sudbury Neutrino Observatory with Enhanced Neutral
Current Sensitivity}, in print and nucl-ex/0309004 
\re
44.\ Kutter T. for SNO Coll.\ 2003, 
{\it Solar Neutrino Results from the Sudbury Neutrino Observatory}, HE 2.2.4
\re
45.\ Bahcall J.N.\ et al.\ 2001, Astrop. J. {\bf 555}, 990 
\re 
46.\ Koshio Y. for Super-Kamiokande Coll.\ 2003 
{\it Recent Results of Solar Neutrino Measurement in 
Super-Kamiokande}, HE 2.2.2 and Suzuki Y., {\it Neutrino Oscillation}, 
plenary talk
\re
47.\ Mitsui T. for KamLAND Coll.
\ 2003 {First Results from KamLAND}, HE 2.2.1 and
Eguchi K.\ et al.\ 2003, Phys. Rev. Lett. {\bf 90}, 021802
\re
48.\ Yoo J. for SK Coll.\ 2003, {\it A Study of Short-Time Periodic 
Variation of $^{8}$B Solar Neutrino Flux at Super-Kamiokande}, HE 2.2.5
\re
49.\ Eguchi K.\ et al.\ 2003, in press and hep-ex/0310047
\re
50.\ Kutter T. for SNO Coll.\ 2003{\it Antineutrino Search at the Sudbury 
Neutrino Observatory}, 1-P-252
\re
51.\ Gando Y.\ 2003, {\it Search for $\bar{\nu}_e$ from the Sun at Super-Kamiokande-I}, HE 2.2.3
\re
52.\ Malek M.S. for Super-Kamiokande Coll.,  
{\it Supernova Relic Neutrino Search Results from Super-Kamiokande}, HE 2.3.1
\re
53.\ Keil M.K., Raffelt G. and Janka H.\ 2002, Astrop. J. {\bf 590}, 971
\re 
54.\ Fulgione W. for LVD Coll.\ 2003, {\it 10 Years Search for Neutrino 
Bursts with LVD}, HE 2.3.9 and
Selvi M. for LVD Coll.\ 2003, {\it Study of the Effect of Neutrino Oscillation
in the Supernova Neutrino Signal with the LVD detector}, 1-P-255
\re
55.\ Namba T. for Super-Kamiokande Coll.\ 2003, {\it Search for Neutrino
Bursts from Supernova Explosions at Super-Kamiokande}, HE 2.3.3
\re
56.\ Feser T. for AMANDA Coll.\ 2003, {\it Online Search for Neutrino Bursts
from Supernovae with the AMANDA Detector}, 1-P-258
\re 
57. Cei F., private communication
\re
58.\ Learned J.G. and Mannheim K.\ 2000, Ann. Rev. Nucl. Part. Sci. 
{\bf 50}, 679 
\re
59.\ Ambrosio M.\ et al.\ 2001 Astrop. J. {\bf 546}, 1038 and
Montaruli T., {\it Neutrino Astrophysics} 
to appear in Proc. of TAUP2003 Conference, Seattle, 5-9 Sep 2003
\re
60.\ Washburn K. for Super-Kamiokande Coll.\ 2003, {\it A Search for Astronomical
Neutrino Sources with the Super-Kamiokande Detector}, HE 2.3.1
\re
61.\ K\"opke L., {\it Recent Results from AMANDA Neutrino Telescope}, 
highlight talk and http://amanda.uci.edu
\re
62.\ Kowalski M.P. for Baikal Coll.\ 2003, {\it Results from the BAIKAL 
Neutrino Telescope}, HE 2.3.11
\re
63.\ Yoshida S. for IceCube Coll \ 2003, {\it The IceCube high energy 
neutrino telescope}, HE 2.3.12
\re
64.\ Montaruli T. for the ANTARES Coll.\ 2003, 
{\it ANTARES Status Report}, 1-P-262 and http://antares.in2p3.fr
\re
65.\ Grieder P.K.F. for NESTOR Coll.\ 2003, {\it NESTOR Neutrino Telescope
Status Report}, 1-P-266
\re
66.\ NEMO-RD Home page: http://nemoweb.lns.infn.it
\re
67.\ Seunarine S. for RICE Coll.\ 2003, 
{\it Updated limits on the Ultra-High Energy 
Neutrino Flux from the RICE Experiment at the South Pole}, HE 2.3.10
\re
68.\ Karle A. for AMANDA Coll.\ 2003, {\it Search for Extraterrestrial Point 
Sources of Neutrinos with AMANDA-II}, HE 2.3.5
\re
69. Ahrens J.\ et al.\ 2003, Astrop. J. {\it 583}, 1040
\re
70.\ Desiati P. for AMANDA Coll.\ 2003, 
{\it Response of AMANDA-II to Cosmic Ray Muons}, 1-P-265
\re
71.\ Geenen H. for AMANDA Coll.\ 2003, {\it Atmospheric Neutrino and Muon Spectra
Measured with the AMANDA-II Detector}, HE 2.3.6 
\re
72.\ Chirkin D.A. for AMANDA Coll.\ 2003, {\it Cosmic Ray Flux Measurement with 
AMANDA-II}, HE 2.1.13
\re
73.\ Gaisser T.K. for IceTop Coll.\ 2003, {\it IceTop: the Surface Component
of IceCube}, HE 1.5.31
\re
74.\ Circella M. for ANTARES Coll.\ 2003, {\it Toward the ANTARES Neutrino 
Telescope: Results from a Prototype Line}, HE 2.5.1
\re
75.\ Desai S.A. for Super-Kamiokande Coll.\ 2003, {\it Study of 
Upward Showering Muons in Super-Kamiokande}, HE 3.3.5
\re
76.\ Heijboer A. for ANTARES Coll.\ 2003, {\it Point Source 
Searches with the ANTARES Neutrino Telescope}, HE 2.3.7
\re
77.\ Hundertmark S. for AMANDA Coll.\ 2003, {\it AMANDA-B10 
Limit on UHE Muon-Neutrinos}, 1-P-256
\re
78.\ Colin Hill G. for AMANDA Coll.\ 2003, {\it Search for Diffuse Fluxes of 
Extraterrestrial Muon-Neutrinos with the AMANDA Detectors}, 1-P-257
\re
79.\ Romeyer A. for ANTARES Coll.\ 2003, {\it Muon Energy Reconstruction in ANTARES
and its Application to the Diffuse Neutrino Flux}, HE 2.3.8
\re
80.\ Kowalski M.P. for AMANDA Coll.\ 2003, {\it Search for High Energy 
Neutrinos of All Flavors with AMANDA-II}, HE 2.3.4
\re
81.\ Yoshida S.\ 2003, {\it Propagation of Extremely High Energy Leptons in 
the Earth}, 1-P-282  
\re
82.\ Dutta S.I., Mocioiu I., Reno M.H. and Sarcevic I.\ 2003,
{\it Tau Neutrinos at EeV Energies}, 1-P-276 
\re
83.\ Athar et al.\ 2003, {\it High Energy Tau Neutrinos: Production, 
Propagation and Prospects of Observations}, HE 2.4.3
\re
84.\ Huang M.A., Tseng J.J. and Lin G.L.\ 2003, 
{\it Energy Fluctuation of Tau Leptons Emerging from Earth}, 1-P-275
\re
85.\ Bugaev E., Montaruli T. and Sokalski I.\ 2003, {\it Detection of
Tau Neutrinos in Underwater Neutrino Telescopes}, 1-P-267 
\re
86.\ Cao Z.\ et al.\ 2003, {Ultra High Energy $\nu_{\tau}$ Detection
Using Air Shower Fluorescence/Cerenkov Light Detector}, HE 2.3.13
\re
87.\ Spergel D.N.\ et al.\ 2003, Astrophys. J. Suppl. {\bf 148}, 175
\re
88.\ Carson M.J. for Boulby Dark Matter Coll.\ 2003, {\it Dark Matter Experiments
at Boulby Mine}, HE 3.3.9
\re
89.\ Morselli A: \ et al.\ 2003, {\it Search for Supersymmetric Dark Matter with 
GLAST}, HE 3.4.3 
\re
90.\ Barwick S.W.\ et al.\ 1997, Astrophys. J. {\bf 482}, 963
\re 
91.\ Lamanna G. for AMS Coll.\ 2003 ,{\it Astroparticle Physics with 
AMS-02}, HE 3.4.1 
\re
92.\ Donato F.\ et al.\ 2003, {\it Cosmic Ray Antiprotons from Relic 
Neutralinos in a Diffusion Model}, 2-P-290 
\re
93.\ Hofmann W. for HEGRA Coll.\ 2003, {\it Search for TeV Gamma-Rays
from the Andromeda Galaxy and for Supersymmetric Dark Matter in the
Core of M31}, HE 3.3.7
\re
94.\ Fleysher L. for Milagro Coll.\ 2003, {\it Search for Relic Neutralinos with 
Milagro}, HE 3.3.3
\re
95.\ Olbrechts P. for AMANDA Coll.\ 2003, 
{\it Search for Muons from WIMP Annihilation in the 
Center of the Earth with the AMANDA-B10 Detector}, HE 3.3.6
\re
96.\ Gondolo P., Edsj\"o J., Bergstr\"om L., Ullio P. and Baltz T.,   
\\
http://www.physto.se/\%7eedsjo/darksusy
\re
97.\ Ambrosio M.\ et al.\ 1999, Phys. Rev. {\bf D60}, 082002
and Montaruli T., {\it Neutrino measurement 
with MACRO: neutrino oscillation, dark matter and astronomy 
studies Les Houches}, Proc. of School and Workshop on Neutrino
Particle Astrophysics, Les Houches, 21 Jan-1 Feb,
available at http://leshouches.in2p3.fr
\re
98.\ Bernabei R.\ et al.\ 2003, {\it Further Results on the 
WIMP Annual Modulation Signature by DAMA/NaI},
to appear in Proc. of TAUP2003, Seattle, 5-9 Sep. 2003 
\re
99.\ Feldman G.J. and Cousins R.D.\ 1998, Phys. Rev. {\bf D57}, 3873
\re
100.\ Waxman E. and Bahcall J.N.\ 1999, Phys. Rev. {\bf D59}, 023002
\re
101.\ Benoit A.\ et al.\ 2002, Phys. Lett. {\bf B545}, 43
\re
102.\ Thompson L. for ANTARES Coll.\ 2003, {\it Dark Matter Searches with the ANTARES
Neutrino Telescope}, HE 3.4.4.
}
\endofpaper
\end{document}